\documentclass[10pt,twocolumn,letterpaper]{article}

\usepackage{iccv}
\usepackage{times}

\usepackage{booktabs}
\usepackage{amsmath}
\usepackage{amssymb}
\usepackage{graphicx}
\usepackage{multirow}
\makeatletter
\@namedef{ver@everyshi.sty}{}
\makeatother

\usepackage{tikz}
\usepackage{tikz}
\newcommand*\emptycirc[1][0.6ex]{\tikz\draw (0,0) circle (#1);} 
\newcommand*\halfcirc[1][0.6ex]{%
	\begin{tikzpicture}
	\draw[fill] (0,0)-- (90:#1) arc (90:270:#1) -- cycle ;
	\draw (0,0) circle (#1);
	\end{tikzpicture}}
\newcommand*\fullcirc[1][0.6ex]{\tikz\fill (0,0) circle (#1);} 

\usepackage{subcaption}
\usepackage{amsfonts}
\usepackage{dsfont}
\usepackage{pifont}
\newcommand{\cmark}{\ding{51}}%
\newcommand{\xmark}{\ding{55}}%

\usepackage[pagebackref=true,breaklinks=true,letterpaper=true,colorlinks,bookmarks=false]{hyperref}

\usepackage[symbol]{footmisc}
\renewcommand{\thefootnote}{\fnsymbol{footnote}}

\definecolor{hero}{rgb}{0.2, 0.2, 0.6}

\iccvfinalcopy 

\def\iccvPaperID{5313} 

\ificcvfinal\fi

\begin{document}
\title{Towards Good Practices in Evaluating Transfer Adversarial Attacks}

\author{Zhengyu Zhao\footnotemark[1]~$^{ ,1}$, \quad Hanwei Zhang\footnotemark[1]~$^{ ,2}$ \quad Renjue Li\footnotemark[1]~$^{ ,3,4}$\\ Ronan Sicre$^2$ \quad  Laurent Amsaleg$^5$  \quad Michael Backes$^1$ \\
$^1$CISPA Helmholtz Center for Information Security \quad\\ $^2$LIS - Ecole Centrale Marseille \quad $^3$SKLCS, Institute of Software, CAS \quad\\$^4$University of Chinese Academy of Sciences \quad $^5$Inria, Univ Rennes, CNRS, IRISA\\
\tt\small \{zhengyu.zhao,director\}@cispa.de, \quad\\
\tt\small \{hanwei.zhang,ronan.sicre\}@lis-lab.fr, lirj19@ios.ac.cn, laurent.amsaleg@irisa.fr
}

\maketitle

\ificcvfinal\thispagestyle{empty}\fi

\begin{abstract}
\footnotetext[1]{Equal contribution}
\renewcommand{\thefootnote}{\arabic{footnote}}

Transfer adversarial attacks raise critical security concerns in real-world, black-box scenarios.
However, the actual progress of this field is difficult to assess due to two common limitations in existing evaluations.
First, different methods are often not systematically and fairly evaluated in a one-to-one comparison.
Second, only transferability is evaluated but another key attack property, stealthiness, is largely overlooked. 
In this work, we design good practices to address these limitations, and we present the first comprehensive evaluation of transfer attacks, covering 23 representative attacks against 9 defenses on ImageNet.
In particular, we propose to categorize existing attacks into five categories, which enables our systematic category-wise analyses.
These analyses lead to new findings that even challenge existing knowledge and also help determine the optimal attack hyperparameters for our attack-wise comprehensive evaluation.
We also pay particular attention to stealthiness, by adopting diverse imperceptibility metrics and looking into new, finer-grained characteristics.
Overall, our new insights into transferability and stealthiness lead to actionable good practices for future evaluations.
Our code and a list of categorized attacks are publicly available at \url{https://github.com/ZhengyuZhao/TransferAttackEval}.

\end{abstract}

\section{Introduction}
\label{sec:intro}
\renewcommand{\thefootnote}{\arabic{footnote}}
Deep Neural Networks (DNNs) have achieved great success in various machine learning tasks.
However, they are known to be vulnerable to adversarial attacks~\cite{goodfellow2014explaining,szegedy2014intriguing}, which intentionally perturb model inputs to induce prediction errors.
An important property of adversarial attacks that makes them threatening in real-world, black-box scenarios is their transferability.


Although extensive studies have been conducted on transfer attacks, we identify two common limitations in existing evaluation practices.
First, existing evaluations are often \emph{unsystematic} and sometimes \emph{unfair}.
Specifically, when a new attack ``A'' is compared to some old attack ``B'', it is often compared in the form of ``A+B vs. B'' instead of ``A vs. B''.
This is especially misleading when both attacks follow similar ideas.
For example, both Admix~\cite{wang2021admix} and SI~\cite{lin2020nesterov} are based on input augmentations, but the original work of Admix only considers ``Admix+SI vs. SI''. 
Moreover, even when ``A vs. B'' is considered for similar attacks, hyperparameter settings are sometimes unfair.
For example, the relatively new input augmentation-based attacks, Admix, SI, and VT~\cite{wang2021enhancing}, adopt multiple input copies by default but are directly compared to earlier attacks, DI~\cite{xie2019improving} and TI~\cite{dong2019evading}, that adopt only one input copy. 


\begin{table*}[!t]
\caption{Overview of our evaluated attacks.}
\newcommand{\tabincell}[2]{\begin{tabular}{@{}#1@{}}#2\end{tabular}}

\renewcommand{\arraystretch}{1}
      \centering
      \resizebox{0.92\textwidth}{!}{
        \begin{tabular}{ccccc}
\toprule[1pt]
\tabincell{c}{Gradient\\Stabilization}&\tabincell{c}{Input\\Augmentation}&\tabincell{c}{Feature\\Disruption}&\tabincell{c}{Surrogate\\Refinement}&\tabincell{c}{Generative\\Modeling}\\
\midrule

&DI~\cite{xie2019improving} (CVPR'19)&TAP~\cite{zhou2018transferable} (ECCV'18)&SGM~\cite{wu2020skip} (ICLR'20)&GAP~\cite{poursaeed2018generative} (CVPR'18)\\
MI~\cite{dong2018boosting} (CVPR'18)&TI~\cite{dong2019evading} (CVPR'19)&AA~\cite{inkawhich2019feature} (CVPR'19)&LinBP~\cite{guo2020backpropagating} (NeurIPS'20)&CDA~\cite{naseer2019cross} (NeurIPS'19)\\
NI~\cite{lin2020nesterov} (ICLR'20)&SI~\cite{lin2020nesterov} (ICLR'20)&ILA~\cite{huang2019enhancing} (ICCV'19)&RFA~\cite{springer2021little} (NeurIPS'21)&TTP~\cite{naseer2021generating} (ICCV'21)\\
PI~\cite{wang2021boosting} (BMVC'21)&VT~\cite{wang2021enhancing} (CVPR'21)&FIA~\cite{wang2021feature} (ICCV'21)&IAA~\cite{zhu2021rethinking} (ICLR'22)&GAPF~\cite{kanth2021learning} (NeurIPS'21)\\
&Admix~\cite{wang2021admix} (ICCV'21)&NAA~\cite{zhang2022improving} (CVPR'22) &DSM~\cite{yang2022boosting} (arXiv'22)&BIA~\cite{zhang2022beyond} (ICLR'22)\\
\bottomrule[1pt]
\end{tabular}
}
\label{tab:attack}
\end{table*}

\begin{table}[!t]
\caption{Overview of our evaluated defenses.}
\newcommand{\tabincell}[2]{\begin{tabular}{@{}#1@{}}#2\end{tabular}}

\renewcommand{\arraystretch}{1}
      \centering
      \resizebox{1\columnwidth}{!}{
        \begin{tabular}{ccc}
\toprule[1pt]

\tabincell{c}{Input\\Pre-processing}&\tabincell{c}{Purification\\Network}&\tabincell{c}{Adversarial\\Training}\\
\midrule
BDR~\cite{xu2017feature} (NDSS'18)&HGD~\cite{liao2018defense} (CVPR'18)&AT$_{\infty}$~\cite{xie2019feature} (CVPR'19)\\
PD~\cite{prakash2018deflecting} (CVPR'18)&NRP~\cite{naseer2020self} (CVPR'20)&FD$_{\infty}$~\cite{xie2019feature} (CVPR'19)\\
R\&P~\cite{xie2017mitigating} (ICLR'18)&DiffPure~\cite{nie2022diffusion} (ICML'22)&AT$_{2}$~\cite{salman2020adversarially} (NeurIPS'20) \\

\bottomrule[1pt]
\end{tabular}}
\label{tab:defense}
\end{table}

Second, existing evaluations mainly focus on transferability but largely overlook another key attack property, \emph{stealthiness}.
Specifically, the stealthiness is mostly addressed by measuring the imperceptibility of image perturbations based on $L_p$ norms.
Using only $L_p$ norms may be sufficient for evaluating white-box attacks but becomes less meaningful when larger perturbations are required~\cite{sharif2018suitability,tramer2020fundamental}, as for our studied transfer attacks.
One recent study~\cite{chen2021measuring} also notices that the transferability of attacks bounded by the same $L_{\infty}$ norm is indeed positively correlated with their $L_{2}$ norms.  
Moreover, other stealthiness measures beyond imperceptibility have not been well explored.

In this work, we present the first systematic evaluation of transfer attacks, covering 23 representative attacks (in 5 categories) against 9 defenses (in 3 categories) on 5000 ImageNet images.
See Table~\ref{tab:attack} and Table~\ref{tab:defense} for an overview of our evaluated attacks and defenses.
In particular, our evaluation addresses the first limitation by proposing a new attack categorization, which enables systematic category-wise analyses.
These analyses lead to new insights that challenge existing knowledge.
For example, we find that using more iterations for gradient stabilization attacks may even decrease the performance.
Another example is that the earliest input augmentation attack, DI, surprisingly outperforms all subsequent attacks when they are compared fairly with the same number of input copies.
These analyses also help determine the optimal hyperparameters for our attack-wise, comprehensive evaluation.
Table~\ref{tab:comp} demonstrates that our evaluation is the most systematic and considers the largest number of attack categories and more complete one-to-one comparisons for both intra-category and inter-category.

We address the second limitation by adopting five diverse imperceptibility metrics beyond the $L_p$ norms and also investigating the finer-grained characteristics of the perturbations and misclassification.
We find that~\emph{all} transfer attacks indeed sacrifice the imperceptibility compared to the basic, PGD attack.
Moreover, the perturbations of different attacks bounded by the same $L_{\infty}$ norm are dramatically different in terms of both the imperceptibility scores and finer-grained stealthiness characteristics.

\begin{table}[!t]
\newcommand{\tabincell}[2]{\begin{tabular}{@{}#1@{}}#2\end{tabular}}
\caption{Comparison methodology in existing work. ``Intra-category'': The new attack A is compared to another attack B in the same category following ``A vs. B''. ``Inter-category'': A is compared to B from another category following ``A vs. B''. The values in brackets report the number of compared attacks.}

\renewcommand{\arraystretch}{1}
      \centering
      \resizebox{0.82\columnwidth}{!}{
        \begin{tabular}{l|ccc}
\toprule[1pt]
Attacks&\tabincell{c}{Total number\\of categories}&\tabincell{c}{Intra-\\category} &\tabincell{c}{Inter-\\category} \\

\midrule
MI~\cite{dong2018boosting}         &1&\xmark&\xmark\\
NI~\cite{lin2020nesterov}          &1&\cmark (1)&\xmark\\
PI~\cite{wang2021boosting}         &1&\cmark (2)&\xmark\\
DI~\cite{xie2019improving}         &2&\xmark&\cmark (1)\\
TI~\cite{dong2019evading}          &2&\xmark&\xmark\\
SI~\cite{lin2020nesterov}          &2&\xmark&\xmark\\
VT~\cite{wang2021enhancing}        &2&\xmark&\xmark\\
Admix~\cite{wang2021admix}         &2&\xmark&\xmark\\
TAP~\cite{zhou2018transferable}    &2&\xmark&\cmark (2)\\
AA~\cite{inkawhich2019feature}     &2&\xmark&\cmark (1)\\
ILA~\cite{huang2019enhancing}      &3&\xmark&\xmark\\
FIA~\cite{wang2021feature}         &3&\cmark (2)&\cmark (3)\\
NAA~\cite{zhang2022improving}      &2&\cmark (3)&\cmark (1)\\
SGM~\cite{wu2020skip}              &3&\xmark&\cmark (3)\\
LinBP~\cite{guo2020backpropagating}&2&\cmark (1)&\cmark (2)\\
RFA~\cite{springer2021little}      &3&\xmark&\xmark\\
IAA~\cite{zhu2021rethinking}       &4&\cmark (1)&\cmark (4)\\
DSM~\cite{yang2022boosting}        &3&\xmark&\xmark\\
GAP~\cite{poursaeed2018generative} &2&\xmark&\cmark (1)\\
CDA~\cite{naseer2019cross}         &4&\cmark (1)&\cmark (7)\\
GAPF~\cite{kanth2021learning}      &2&\cmark (2)&\cmark (1)\\
BIA~\cite{zhang2022beyond}         &3&\cmark (1)&\cmark (3)\\
TTP~\cite{naseer2021generating}    &5&\cmark (2)&\cmark (8)\\
Ours&\textbf{5}&\cmark (\textbf{4})&\cmark (\textbf{22})\\
\bottomrule[1pt]
\end{tabular}
}
\label{tab:comp}
\end{table}

Our evaluation also leads to new general insights.
For example, we find that transferability is highly contextual, i.e., the optimal performance of one (category of) attack/defense against one (category of) defense/attack may not generalize well to another.
Moreover, defenses may largely overfit to specific attacks that are similar to those used in defense optimization.
In particular, we identify that DiffPure~\cite{nie2022diffusion}, which originally claims the start-of-the-art white-box robustness, is surprisingly vulnerable to (black-box) transfer attacks that produce smooth perturbations.

Overall, based on our new insights, we recommend the following actionable good evaluation practices:
\vspace{-0.5em}
\begin{itemize}
\setlength\itemsep{-0.3em}

\item Generate a transferability vs. iteration curve.
\item Set fair hyperparameters for similar attacks.
\item Measure stealthiness beyond a single $L_p$ norm.
\item Test attacks/defenses against diverse defenses/attacks.

\end{itemize}
\section{New Attack Categorization}
\label{sec:cate-trans-attack}

In this section, we categorize transfer attacks into five categories: gradient stabilization, input augmentation, feature disruption, surrogate refinement, and generative modeling.
These five categories are \emph{disjoint}, i.e., each attack falls into \emph{only one} category according to its core component.
In this way, attacks that share the same working mechanism can be systematically compared along a single dimension.
Our categorization is also validated to be reasonable through a perturbation classification task (see Figure~\ref{fig:confusion}).

\subsection{Gradient Stabilization Attacks}
Different DNN architectures tend to yield radically different decision boundaries, yet similar test accuracy, due to their high non-linearity~\cite{liu2017delving,somepalli2022can}.
For this reason, the attack gradients calculated on a specific model may cause the adversarial images to trap into local optima, resulting in low transferability to another, unseen model.
To address this issue, several studies adopt popular machine learning techniques for stabilizing iterative gradient updates.
Specifically, a momentum term can be integrated into the attack optimization to accumulate previous gradients~\cite{dong2018boosting}, and a Nesterov accelerated gradient (NAG) term can be integrated to look ahead~\cite{lin2020nesterov}.
The property of NAG is later examined in~\cite{wang2021boosting}, where using only the last gradient is found to be better than using all previous gradients.

\subsection{Input Augmentation Attacks}
Learning adversarial images that can transfer to unseen models follows a similar principle to learning models that can generalize to unseen test images~\cite{liang2021uncovering}. 
For this reason data augmentation, as a common technique for improving model generalizability, is exploited to improve attack transferability.
To this end, several studies force the adversarial effects to be invariant to
certain semantic-preserving image transformations, such as geometric transformations (e.g., resizing \& padding~\cite{xie2019improving} and translation~\cite{dong2019evading}), pixel value scaling~\cite{lin2020nesterov}, random noise~\cite{wang2021enhancing}, and mixed images from other classes~\cite{wang2021admix}.
Several recent studies also explore more complex methods that refine the above transformations~\cite{zou2020improving,yuan2022adaptive}, leverage regional (object) information~\cite{li2020regional,byun2022improving,wei2022incorporating}, or rely on frequency-domain image transformations~\cite{long2022frequency}.

\subsection{Feature Disruption Attacks}
The cross-entropy loss is commonly adopted because attacks aim at output-level, misclassification.
However, the output-level information is often model-specific.
In contrast, features extracted from DNN intermediate layers are known to be more generic~\cite{yosinski2014transferable,kornblith2019similarity}.
Inspired by this fact, several studies propose to improve attack transferability based on feature disruptions.
The general objective is to modify the image such that its feature is pushed away from the original feature. 
Specifically, early attacks treat all features indiscriminately~\cite{naseer2018task,zhou2018transferable,ganeshan2019fda,huang2019enhancing,liu2019s,li2020yet}, which is shown to cause sub-optimal attack transferability because the model decision may only depend on a small set of important features~\cite{zhou2016learning}.
For this reason, later attacks~\cite{wu2020boosting,wang2021feature,zhang2022improving} instead calculate the feature distance only on important features that are computed based on model interpretability techniques~\cite{selvaraju2017grad,sundararajan2017axiomatic}.
Feature disruption is also explored for targeted attacks, where the image is modified such that its feature becomes more similar to a target image~\cite{inkawhich2019feature} or image distribution~\cite{inkawhich2020transferable,inkawhich2020perturbing}.

\subsection{Surrogate Refinement Attacks}
The current CNN models are commonly optimized towards high prediction accuracy but with much less attention given to transferable representations.
Several recent studies find that adversarial training can help pre-trained models better transfer to downstream tasks, although it inevitably trades off model accuracy in the source domain~\cite{salman2020adversarially,deng2021adversarial,utrera2021adversarially}.
This finding encourages researchers to explore how the attack transferability can be improved based on refining the surrogate model in diverse aspects, such as training procedures~\cite{springer2021little,zhang2021early,yang2022boosting}, local architectures~\cite{wu2020skip,zhu2021rethinking}, and activation functions~\cite{guo2020backpropagating,zhang2021backpropagating,zhu2021rethinking}.
This line of research also leads to rethinking the relation between the accuracy of the surrogate model and the transferability~\cite{zhang2021early,zhu2021rethinking}.

\subsection{Generative Modeling Attacks}
In addition to the above iterative attacks, existing work also uses generative models to improve transferability.
Basically, an image generator is learned on additional data such that it can take as input any original image and outputs an adversarial image with only one forward pass.
During training, the generator is optimized to fool a discriminator, which is a pre-trained fixed classifier.
Specifically, the earliest generative attack~\cite{poursaeed2018generative} adopts the widely-used cross-entropy loss.
Later studies improve it by adopting a relativistic cross-entropy loss~\cite{naseer2019cross} or intermediate-level, feature losses~\cite{kanth2021learning,zhang2022beyond}.
In particular, to improve targeted transferability, class-specific~\cite{naseer2021generating} and class-conditional~\cite{yangx2022boosting} generators are also explored.
In general, a clipping operation is applied during training to constrain the perturbation size.

\section{Evaluation Methodology}
\label{sec:eva-met}
In general, we make sure our evaluation follows the most common settings in existing work regarding five aspects (see Table~\ref{tab:circle} in Appendix~\ref{app:comp} for details).
\subsection{Threat Model}
\label{sec:thr}
We specify our threat model from the following three common dimensions~\cite{papernot2016limitations,biggio2018wild,carlini2019evaluating}.

\noindent\textbf{Adversary's knowledge.} An adversary can have various levels of knowledge about the target model.
In the ideal, \ie white-box case, an adversary has full control over the target model.
In the realistic, \ie black-box case, an adversary has either query access to the target model or in our transfer setting, no access but only leverages a surrogate model.
In this work, we adopt the most common transfer setting in which the surrogate and target models are trained on the same (public) dataset.
Cross-dataset transferability is rarely explored~\cite{naseer2019cross,zhang2022beyond} and beyond the scope of this work.

\noindent\textbf{Adversary's goal.} An adversary aims at either untargeted or targeted misclassification.
An untargeted attack aims to fool the classifier into predicting any other class than the original one, i.e., $f(\boldsymbol{x}')\neq{y}$.
A targeted attack aims at a specific incorrect class $t$, i.e., $f(\boldsymbol{x}')=t$.
In this work, we evaluate the untargeted transferability because most of our considered attacks cannot achieve substantial targeted transferability~\cite{naseer2019cross,naseer2021generating,zhao2021success}.

\noindent\textbf{Adversary's capability.} In practice, an adversary should be constrained to stay stealthy.
Most existing work addresses this by pursuing the imperceptibility of perturbations, based on $L_p$ norms~\cite{carlini2017towards,goodfellow2014explaining,kurakin2016adversarial,szegedy2014intriguing,papernot2016limitations} or other metrics~\cite{rozsa2016adversarial,Croce_2019_ICCV,luo2018towards,zhang2020smooth,xiao2018spatially,kanbak2018geometric,alaifari2018adef,wong2019wasserstein,zhao2020towards}.
There are also recent studies on ``perceptible yet stealthy'' attacks~\cite{bhattad2020Unrestricted,shamsabadi2020colorfool,zhao2020adversarial_arxiv}.
In this work, we generate $L_{\infty}$-bounded perturbations, following the transfer literature.


\subsection{Evaluation Metrics}
\noindent\textbf{Transferability.} The transferability is measured by the (untargeted) success rate. 
Given an attack $\mathcal{A}$ that generates an adversarial image $\boldsymbol{x}'_i$ for its original image with the true label $y_i$ and target classifier $f$, the success rate over $N$ test images is defined as:
\begin{gather}\label{eq:suc}
     \mathrm{Suc}(\mathcal{A})=\frac{1}{N}\sum_{i=1}^{N}\mathbf{1}\big(f(\boldsymbol{x}'_i)\neq y_i\big),
\end{gather}
where $\mathbf{1}(\cdot)$ is the indicator function.

\noindent\textbf{Stealthiness.} The imperceptibility is measured by a variety of metrics: Root Mean Squared Error (RMSE), Peak Signal-to-Noise Ratio (PSNR), Structural Similarity Index Measure (SSIM)~\cite{wang2004image}, $\Delta E$~\cite{luo2001development,zhao2020towards}, Learned Perceptual Image Patch Similarity (LPIPS)~\cite{zhang2018unreasonable}, and Frechet Inception Distance (FID)~\cite{szegedy2016rethinking}.
Detailed definitions of them can be found in Appendix~\ref{app:metric}.
Beyond imperceptibility, we also look into other, finer-grained stealthiness characteristics regarding the perturbation and misclassification (see Section~\ref{sec:eva-res-stea} for details).

\subsection{Experimental Setups}
\noindent\textbf{Attacks and defenses.} For each of the five attack categories presented in Section~\ref{sec:cate-trans-attack}, we select 5 representative attacks (but only 3 for gradient stabilization), resulting in a total number of 23 attacks, as summarized in Table~\ref{tab:attack}.
Following the common practice, the $L_{\infty}$ norm bound is set to $\epsilon=16/255$.
In addition to testing against standard models, we also consider 9 representative defenses from three different categories, as summarized in Table~\ref{tab:defense}.
Detailed descriptions and hyperparameter settings of the selected attacks and defenses can be found in Appendix~\ref{app:att_def}.

\noindent\textbf{Data and models.} 
We focus on the complex dataset, ImageNet, because attack transferability on simple datasets (e.g., MNIST and CIFAR) has been well solved~\cite{inkawhich2019feature}.
For stealthiness, it also makes more sense to consider larger images.
We consider four model architectures, i.e., InceptionV3~\cite{szegedy2016rethinking}, ResNet50~\cite{he2016deep}, DenseNet121~\cite{huang2017densely}, and VGGNet19~\cite{simonyan2015very}.
We randomly select 5000 images (5 per class\footnote{Several classes contain fewer than 5 eligible images.}) from the validation set that are correctly classified by all the above four models.
All original images are resized and then cropped to the size of 299$\times$299 for Inception-V3 and 224$\times$224 for the other four models. 

\section{Analyses in Each Attack Category}
\label{sec:ana}

In this section, we conduct systematic analyses for each of the five attack categories.
In each category, similar attacks are systematically compared along a single dimension with hyperparameters fairly set.
In particular, each attack method is implemented with only its core algorithm, although it may integrate other algorithms by default in its original work.
These analyses also help us determine the optimal hyperparameter settings required for our later comprehensive evaluation in Section~\ref{sec:eva-res-trans}.

\begin{figure}[!t]
\centering
\includegraphics[width=\columnwidth]{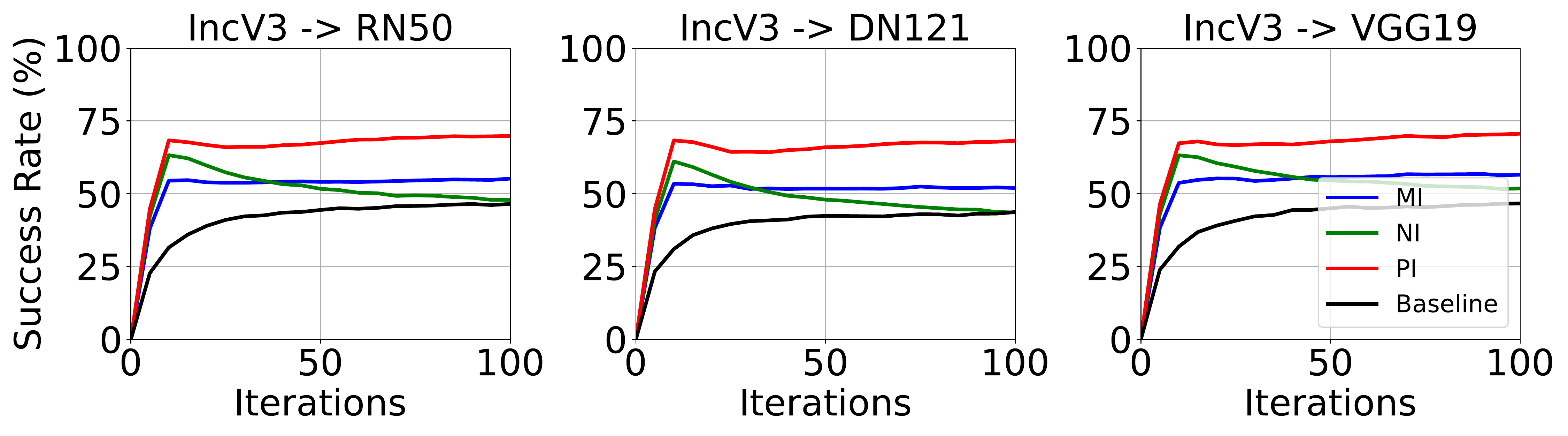}
\includegraphics[width=\columnwidth]{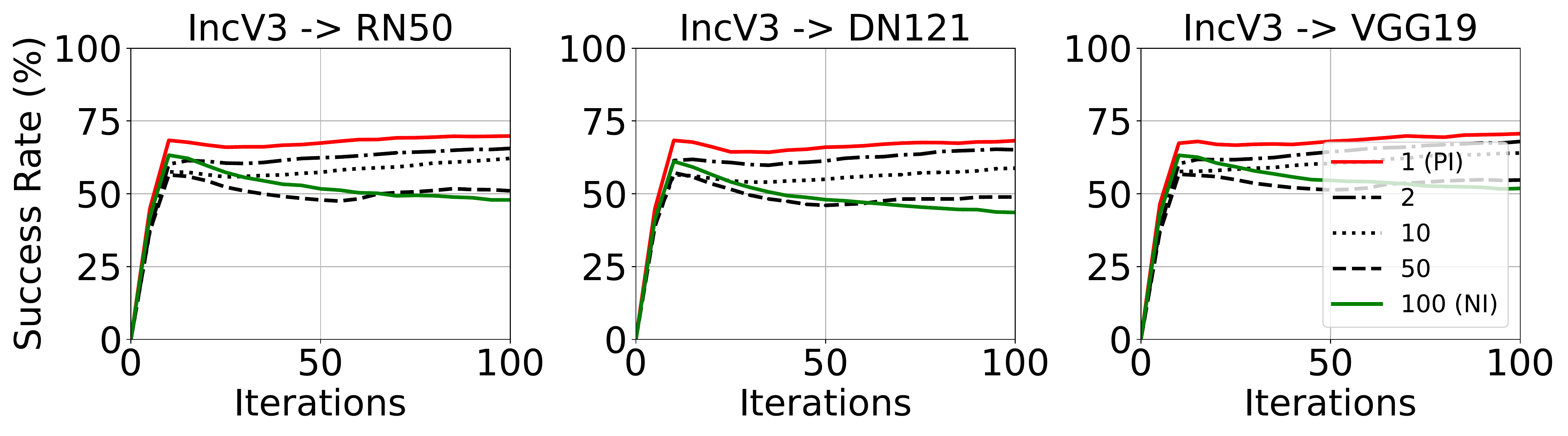}
\caption{\textbf{Results for gradient stabilization attacks.} \textbf{Top:} Transferability vs. iteration curve.
\textbf{Bottom:} Influence of the number of look-ahead iterations.} 
\label{fig:gradient}
\end{figure}

\subsection{Analysis of Gradient Stabilization Attacks}
\label{sec:attack-gs}
Figure~\ref{fig:gradient} Top shows the transferability of the three gradient stabilization attacks under various iterations.
As can be seen, all three attacks converge very fast, within 10 iterations, since they accumulate gradients with the momentum.
However, using more iterations does not improve and may even harm the performance.
This finding indicates that in practice, we should early stop such attacks in order to ensure optimal transferability.
More specifically, PI and NI perform better than MI due to the use of looking ahead, but NI is better only at the beginning.

The only difference between PI and NI is the maximum number of previous iterations used to look ahead.
To figure out the impact of the number of look-ahead iterations, we repeat the experiments with values from 1 (i.e., PI) to 100 (i.e., NI).
Figure~\ref{fig:gradient} Bottom shows that the performance drops more as more previous iterations are incorporated to look ahead, and the optimal performance is actually achieved by PI.

\begin{figure}[!t]
\centering
\includegraphics[width=\columnwidth]{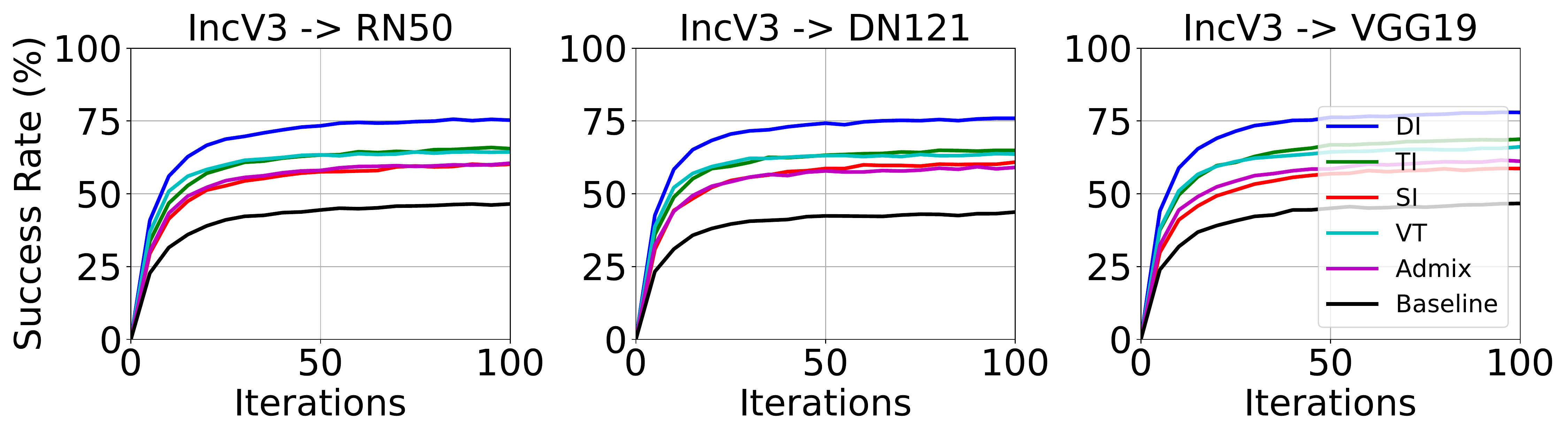}
\includegraphics[width=\columnwidth]{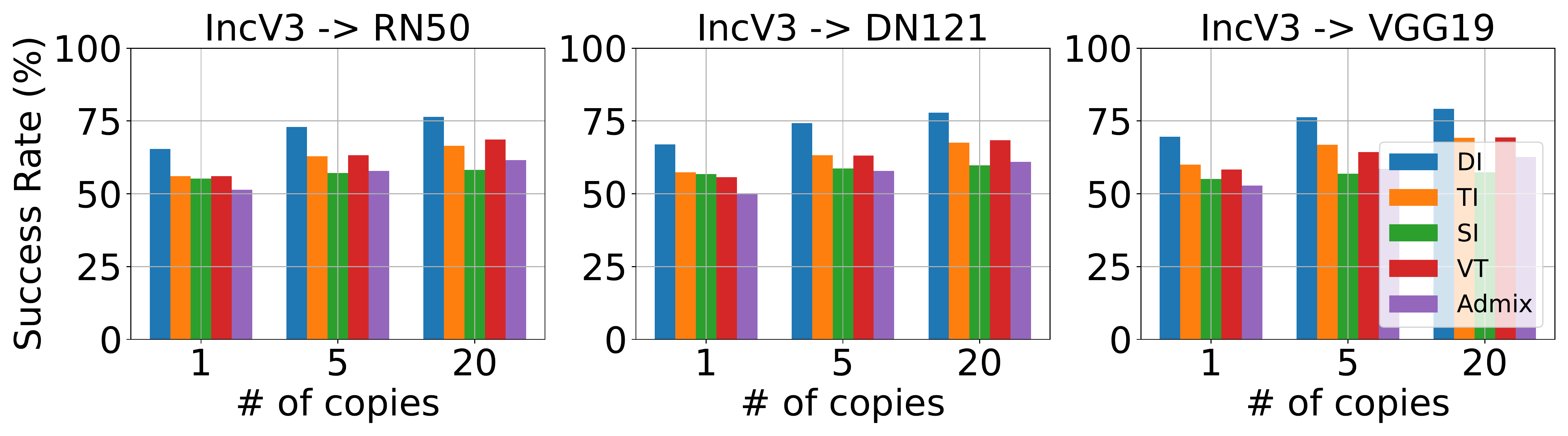}
\caption{\textbf{Results for input augmentation attacks.} \textbf{Top:} Transferability vs. iteration curve. All attacks use five input copies. \textbf{Bottom:} Impact of the number of input copies.}
\label{fig:copy_5}
\end{figure}

\begin{figure}[!t]
\centering
\includegraphics[width=\columnwidth]{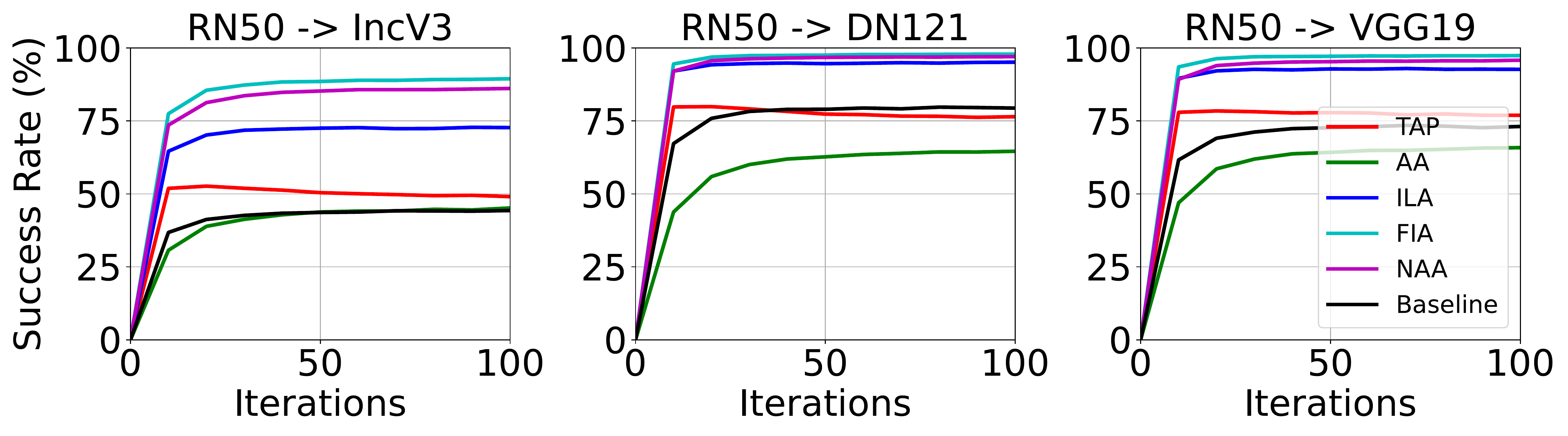}
\includegraphics[width=\columnwidth]{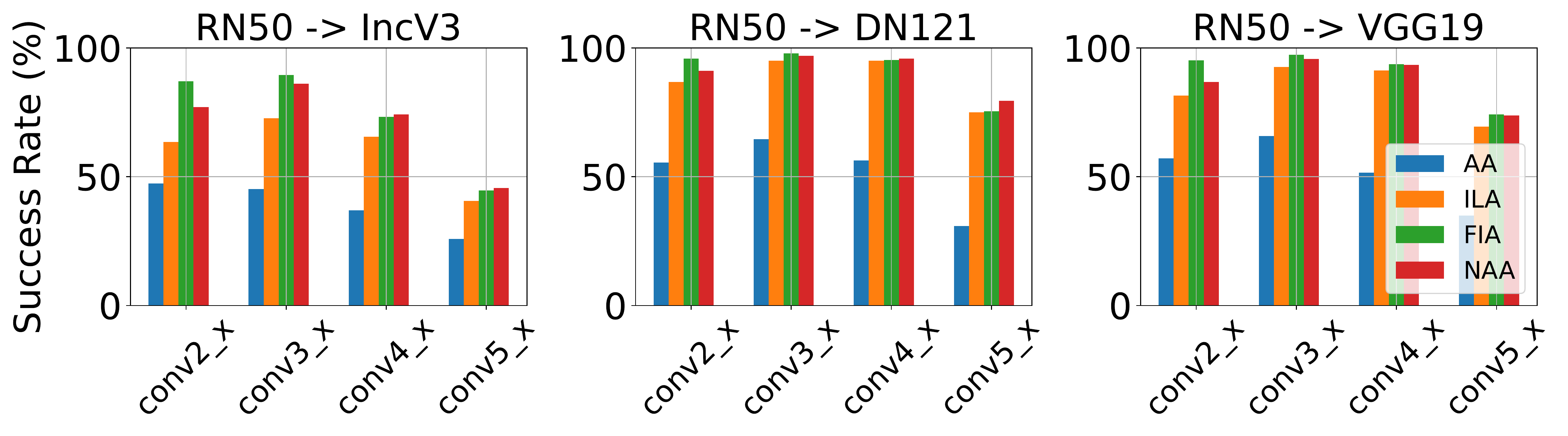}
\caption{\textbf{Results for feature disruption attacks.} \textbf{Top:} Transferability vs. iteration curve. TAP uses all layers and the others use ``conv3\_x''. \textbf{Bottom:} Impact of the layers.}
\label{fig:intermediate_resnet50}
\end{figure}

\subsection{Analysis of Input Augmentation Attacks}
\label{sec:attack-ia}
Figure~\ref{fig:copy_5} Top shows the transferability of the five input augmentation attacks for various iterations.
Previous evaluations of input augmentation attacks often conduct unfair comparisons because different attacks may leverage a different number of random input copies~\cite{wang2021admix}.
Differently, we compare all attacks with the same number of random input copies. 
Surprisingly, we find that the earliest method, DI, always performs the best, and another early method, TI, often performs the second best.
In contrast, the latest method, Admix, achieves very low transferability. 

We explain this new observation by comparing the input diversity caused by different attacks since higher input diversity generally leads to higher transferability~\cite{xie2019improving,long2022frequency}.
To this end, we quantify the input diversity of the five attacks by the impact of their augmentations on the model's top-1 logit value, over 5000 images with 10 repeated runs. 
We get 0.23 for DI, 0.0017 for TI, 0.00039 for SI, 0.000091 for VT, and -0.035 for Admix.
These results confirm the highest input diversity of DI and the lowest of Admix.
In particular, Admix is not a good label-preserving augmentation since it even decreases the logit.

We further explore the impact of the number of random input copies.
In Figure~\ref{fig:copy_5} Bottom, we see that the transferability of all attacks is improved when more copies are used.
In addition, the superiority of DI and TI consistently holds in all settings.
This suggests the superiority of spatial transformations (i.e. resizing\&padding in DI and translation in TI) over other transformations, such as pixel scaling in SI, additive noise in VT, and image composition in Admix.
Note that more copies generally consume  more computational resources.

\begin{figure}[!t]
\centering
\includegraphics[width=\columnwidth]{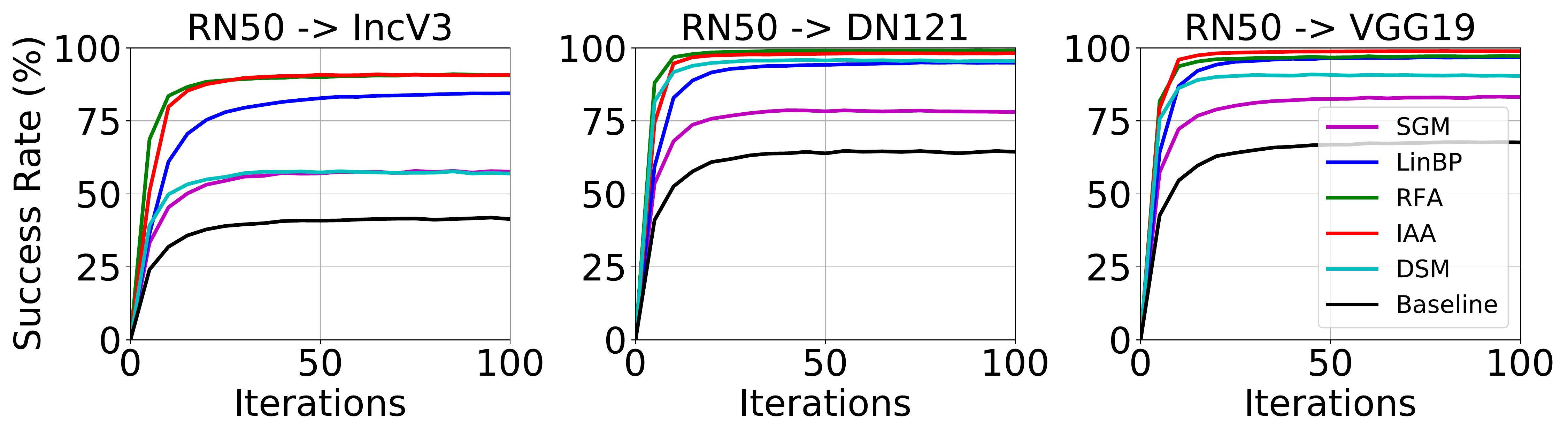}
\caption{\textbf{Results for surrogate refinement attacks (1/2):} Transferability vs. iteration curve.
}
\label{fig:surrogateR_resnet50}
\end{figure}

\begin{table}[!t]
\caption{\textbf{Results for surrogate refinement attacks (2/2):} Impact of the surrogate property. Definitions of the last three metrics can be found in Appendix~\ref{app:inter}.}
\newcommand{\tabincell}[2]{\begin{tabular}{@{}#1@{}}#2\end{tabular}}

\renewcommand{\arraystretch}{1}
      \centering
      \resizebox{0.85\columnwidth}{!}{
        \begin{tabular}{l|c|cccc}
\toprule[1pt]
    Attacks   &  Transfer$\uparrow$ &Acc$\uparrow$  &AI$\uparrow$ &AD$\downarrow$      &KL$\downarrow$\\
\midrule

SGM      &90.3&\textbf{100.0} &37.4&8.4  &\textbf{14.8}\\
LinBP    &98.1&\textbf{100.0}  &37.4&8.4  &\textbf{14.8}\\
RFA      &99.1&96.4 &33.8 &14.2  &16.0\\
IAA      & \textbf{99.5}  &34.9   &4.4&29.5  &22.4\\
DSM      &95.7 &97.6  &\textbf{41.0}&\textbf{7.6}  &15.9\\

\bottomrule[1pt]
\end{tabular}
}
\label{tab:surrogateR_resnet50}
\end{table}

\subsection{Analysis of Feature Disruption Attacks}
Figure~\ref{fig:intermediate_resnet50} Top shows the transferability of the five feature disruption attacks for various iterations.
We see that FIA and NAA, which exploit feature importance, achieve the best results.
In addition, ILA and TAP perform well, by incorporating the CE loss into their feature-level optimizations.
Finally, AA performs worse than the baseline since it only uses a feature loss.

Since all these attacks, except TAP, only disrupt features in a specific layer, we explore the impact of layer choice.
In Figure~\ref{fig:intermediate_resnet50} Bottom, we see that the last layer (``conv5\_x'') performs much worse than the early layers.
This might come from the last layer being too complex and model-specific, which results in poor generalizability to unseen models.
Moreover, the mid layer (``conv3\_x'') always achieves the best performance since it can learn more semantic features but earlier layers normally capture simple features, e.g., colors and textures~\cite{zeiler2014visualizing}.

\subsection{Analysis of Surrogate Refinement Attacks}
Figure~\ref{fig:surrogateR_resnet50} shows the transferability of the five surrogate refinement attacks for various iterations.
Here ResNet50 is used as the surrogate model for a fair comparison since SGM~\cite{wu2020skip} can only be applied to architectures with skip connections.
We see that IAA achieves the best results since it optimizes the hyperparameters of skip connections and continuous activation functions.
In addition, RFA achieves much lower performance (sometimes even lower than the baseline attack) since the standard target and robust surrogate models may rely on distinct features~\cite{ilyas2019adversarial}.

Model refinement opens a direct way to explore the impact of specific model properties of the surrogate on transferability.
Thus we look into model properties in terms of two common metrics: accuracy and interpretability, as well as model similarity to the target model, DenseNet121.
Specifically, interpretability is measured by Average Increase (AI) and Average Drop (AD)~\cite{chattopadhay2018grad} based on GradCAM~\cite{selvaraju2017grad}, and model similarity is measured based on the Kullback-Leibler (KL) divergence of output features following~\cite{inkawhich2020transferable}.
Note that we additionally refine the surrogate models in LinBP and SGM during the forward pass following the same hyperparameters used in backpropagation.
Table~\ref{tab:surrogateR_resnet50} shows a clear negative correlation between transferability and other model properties.
Specifically, IAA achieves the highest transferability but the worst performance regarding all the other four metrics.

\subsection{Analysis of Generative Modeling Attacks}
\begin{figure}[!t]
\centering
\includegraphics[width=\columnwidth]{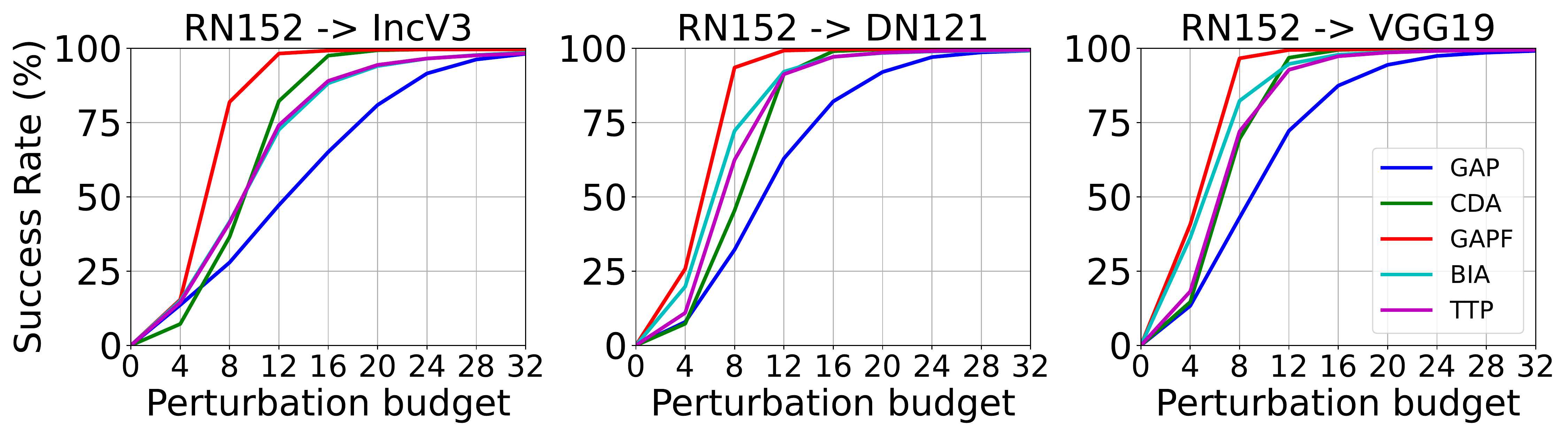}
\includegraphics[width=\columnwidth]{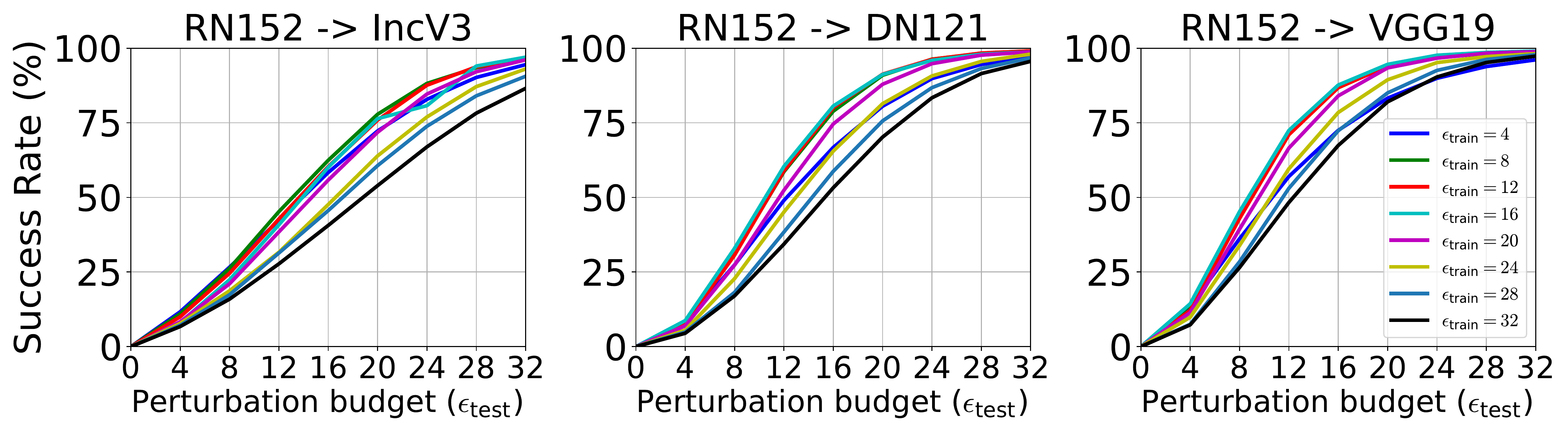}
\caption{\textbf{Results for generative modeling attacks.} \textbf{Top:} Transferability vs. perturbation budget curve. \textbf{Bottom:} Impact of the training perturbation bound, for GAP attack.}
\label{fig:gen_test}
\end{figure}

Generative modeling attacks can generate perturbations for any given image with only one forward pass.
These perturbations output from the generator are initially unbounded and then clipped to satisfy the imperceptibility.
This means that once trained, the generator can be used to generate perturbations under various constraints.
Figure~\ref{fig:gen_test} Top shows the transferability of the five generative modeling attacks under various perturbation bounds.
For the targeted attack TTP, we calculate its untargeted transferability over 5000 targeted adversarial images that are generated following the 10-Targets setting~\cite{naseer2021generating}.
As can be seen, the transferability generally increases as the perturbation constraint is relaxed.
Consistent with our discussion about feature disruption attacks, feature-level losses (i.e., GAPF and BIA) outperform output-level losses (i.e., GAP and CDA).
One exception is TTP, which is not good for small norms because it relies on target semantics~\cite{naseer2021generating,zhao2021success}.

When training the generator, existing work commonly adopts the perturbation bound $\epsilon_{\textrm{train}}=10$.
For this reason, it is worth exploring the impact of this training perturbation bound on attack transferability.
Thus, we train the GAP generator with various $\epsilon_{\textrm{train}}$ and test these generators across different perturbation bounds $\epsilon_{\textrm{test}}$.
Figure~\ref{fig:gen_test} Bottom shows that adjusting $\epsilon_{\textrm{train}}$ has a substantial impact in general.
Another unexpected finding is that using the same bound for training and testing (i.e., $\epsilon_{\textrm{train}}=\epsilon_{\textrm{test}}$) does not lead to better results, especially when the bound is large.
Instead, using a moderate $\epsilon_{\textrm{train}}$ (i.e. 8-16) leads to the best results, which also suggests that the commonly used 10 in existing work is indeed optimal.

\begin{table*}[!t]
\caption{Attack transferability in terms of success rate (\%). ResNet-50 is used as the surrogate.}
\newcommand{\tabincell}[2]{\begin{tabular}{@{}#1@{}}#2\end{tabular}}

\renewcommand{\arraystretch}{1}
      \centering
      \resizebox{0.95\textwidth}{!}{
        \begin{tabular}{l|cccc|ccccccccc}
\toprule[1pt]
\multirow{2}{*}{Attacks}&\multicolumn{4}{c|}{Without Defenses}&\multicolumn{3}{c}{Input Pre-processing}&\multicolumn{3}{c}{Purification Network}&\multicolumn{3}{c}{Adversarial Training}\\
&IncV3&DN121&VGG19&ViT&BDR&PD&R\&P&HGD&NRP&DiffPure&AT$_{\infty}$&FD$_{\infty}$&AT$_{2}$\\
\midrule

Clean Acc&100.0&100.0&100.0&94.7&89.1&97.3&94.1&98.0&90.2&91.7&77.8&81.4&43.2\\
PGD&43.6&79.6&72.7&21.2&100.0&100.0&98.2&84.5&78.9&13.1&22.1&18.6&61.2\\
\hline
MI~\cite{dong2018boosting}&55.8&85.7&78.1&31.4&100.0&100.0&98.6&87.5&49.4&21.4&22.6&18.9&64.9\\
NI~\cite{lin2020nesterov} &60.4&87.2&82.7&31.1&100.0&100.0&99.2&88.2&\underline{83.3}&20.0&22.4&19.1&64.3\\
PI~\cite{wang2021boosting}&66.0&92.1&87.7&33.3&100.0&100.0&99.7&81.3&\textbf{93.3}&20.4&22.8&19.2&64.4\\

\hline
DI~\cite{xie2019improving} &69.8&99.0&99.1&36.9&100.0&100.0&100.0&98.9&81.4&16.1&22.9&19.5&62.3\\
TI~\cite{dong2019evading}  &63.0&96.9&96.1&34.5&100.0&100.0&100.0&97.9&78.4&16.1&22.9&19.1&62.3\\
SI~\cite{lin2020nesterov}  &61.8&93.8&85.6&29.5&100.0&100.0&99.2&95.2&72.6&14.6&22.6&19.2&62.6\\
VT~\cite{wang2021enhancing}&67.1&95.4&92.5&35.9&100.0&100.0&99.8&97.8&80.5&19.1&22.8&19.1&63.3\\
Admix~\cite{wang2021admix} &53.4&86.7&83.5&26.5&100.0&95.4&95.0&89.9&80.0&14.1&22.7&19.0&62.1\\
\hline
TAP~\cite{zhou2018transferable} &50.3&77.0&77.6&28.0&100.0&100.0&95.6&80.5&40.9&15.8&22.6&19.4&64.4\\
AA~\cite{inkawhich2019feature}  &43.5&61.8&64.1&26.1&91.0&88.4&81.4&67.8&14.7&20.6&24.9&21.4&65.1\\
ILA~\cite{huang2019enhancing}   &72.3&94.5&92.6&38.3&100.0&100.0&99.1&94.8&\underline{83.5}&18.2&22.6&19.1&63.9\\
FIA~\cite{wang2021feature}      &88.4&97.5&97.1&61.8&100.0&99.9&99.6&98.1&71.6&28.5&25.1&21.3&68.2\\
NAA~\cite{zhang2022improving}   &85.0&96.7&95.3&55.0&99.9&99.9&99.1&97.2&76.8&29.2&24.8&21.3&66.2\\
\hline
SGM~\cite{wu2020skip}               &57.0&90.3&87.1 &31.8  &90.4&100.0&97.6&94.5&83.1&14.9&22.7&18.9&62.2\\
LinBP~\cite{guo2020backpropagating} &82.8&98.1&97.4&40.0&96.6&100.0&99.2&98.3&66.8&17.7&23.1&19.4&63.8\\
RFA$_2$~\cite{springer2021little}       &89.2 &99.1&96.7&63.1&98.9&99.5&97.9&99.4&34.8&\underline{43.8}&24.8&21.2&67.6\\
RFA$_{\infty}$~\cite{springer2021little}     &66.0&77.7&68.7&\underline{73.3}&81.7&80.4&77.5&72.8&21.0&\textbf{69.4}&\textbf{62.0}&\textbf{53.4}&\textbf{87.3}\\
IAA~\cite{zhu2021rethinking}        &90.8&\underline{99.5}&\underline{99.2}&50.2&98.9&100.0&99.9&\textbf{99.9}&65.4&20.6&23.0&19.5&65.5\\
DSM~\cite{yang2022boosting}         &57.3&95.7&90.8&27.3&97.4&99.1&92.6&98.1&38.2&14.9&22.7&19.1&62.3\\
\hline
GAP~\cite{poursaeed2018generative} &65.1&82.1&87.4&34.8&85.1&87.0&83.5&93.9&2.0&16.8&21.7&18.2&64.2\\
CDA~\cite{naseer2019cross}         &\underline{97.8}&\underline{99.2}&\underline{99.2}&\textbf{82.1}&97.6&100.0&98.3&\underline{99.9}&3.9&\underline{62.0}&\underline{26.0}&\underline{22.6}&\underline{73.2}\\
GAPF~\cite{kanth2021learning}      &\textbf{99.2}&\textbf{99.6}&\textbf{99.6}&\underline{72.5}&99.5&99.6&99.8&\underline{99.8}&14.3&28.9&23.4&20.2&67.7\\
BIA~\cite{zhang2022beyond}         &88.2&97.1&97.8&50.3&96.9&95.4&96.6&98.7&2.6&20.0&\underline{25.3}&\underline{22.2}&66.7\\
TTP~\cite{naseer2021generating}    &89.0&97.1&97.3&64.3&97.6&98.2&97.9&95.4&36.2&37.4&\underline{25.3}&21.3&\underline{68.7}\\

\bottomrule[1pt]
\end{tabular}
}
\label{tab:transfer}
\end{table*}

\section{Comprehensive Evaluation Results}
\label{sec:eva-res-trans}

In this section, we evaluate all 23 attacks on their transferability against both standard and defended target models, as well as stealthiness regarding diverse metrics.
Our comprehensive analyses in Section~\ref{sec:ana} provide evidence for category-level hyperparameter settings of different attacks.
Specifically, for gradient stabilization attacks, we set the iteration number to 10 for early stopping, while for the other three categories of iterative attacks, we set the iteration number to 50 to ensure attack convergence.
For input augmentation attacks, we set the number to 5 to offer a good trade-off between performance and computations.

\subsection{Transferability Results}
Table~\ref{tab:transfer} summarizes the evaluation results in both the standard transfer and defense settings.
Here, in addition to CNN architectures, we also consider ViT-B-16-224~\cite{dosovitskiy2020image} as the target model.
In general, we find that the optimal performance of one (category of) attack/defense against one (category of) defense/attack may not generalize well to another.
More specifically, we make the following observations.

From the attack perspective: \textbf{New model design largely increases attack success.}
We find that in almost all cases, the generative modeling and surrogate refinement attacks, which design new surrogate or generative models, achieve better performance than the other three categories of attacks, which instead use off-the-shelf surrogate models.
\textbf{Transferring from ResNet50 to Inception-v3 is harder than to other architectures.} This is consistent with previous findings in~\cite{inkawhich2020transferable,inkawhich2020perturbing,zhao2021success} and might be due to the fact that the Inception architecture contains relatively complex components, e.g., multiple-size convolution and two auxiliary classifiers.

\begin{figure}[!t]
\centering
\begin{subfigure}[b]{0.15\columnwidth}
\includegraphics[width=\columnwidth]{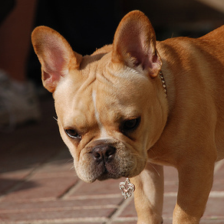}
\caption*{Original}
\end{subfigure}
\hspace{-0.15cm}
\begin{subfigure}[b]{0.15\columnwidth}
\includegraphics[width=\columnwidth]{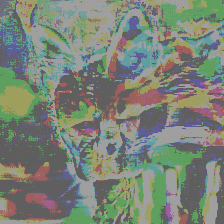}
\caption*{RFA$_{\infty}$}
\end{subfigure}
\hspace{-0.15cm}
\begin{subfigure}[b]{0.15\columnwidth}
\includegraphics[width=\columnwidth]{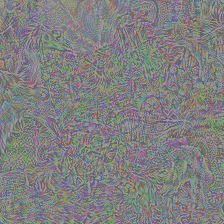}
\caption*{DI}
\end{subfigure}
\hspace{0.1cm}
\begin{subfigure}[b]{0.15\columnwidth}
\includegraphics[width=\columnwidth]{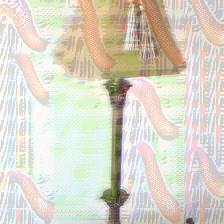}
\caption*{CDA}
\end{subfigure}
\hspace{-0.15cm}
\begin{subfigure}[b]{0.15\columnwidth}
\includegraphics[width=\columnwidth]{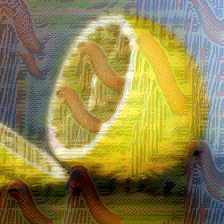}
\caption*{CDA}
\end{subfigure}
\hspace{-0.15cm}
\begin{subfigure}[b]{0.15\columnwidth}
\includegraphics[width=\columnwidth]{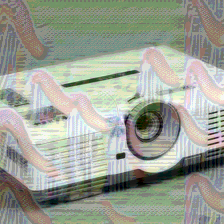}
\caption*{CDA}
\end{subfigure}

\caption{Perturbation Visualizations. \textbf{Left:} The original image along with semantics-aligned, RFA$_{\infty}$ vs. noisy, DI perturbations. \textbf{Right:} CDA yields similar perturbations for different images. Visualizations of all attacks can be found in our GitHub repository.}
\label{fig:vis}
\end{figure}

From the defense perspective: \textbf{Adversarial training is generally effective but input pre-processing is not.} Both the two $L_{\infty}$ adversarial training defenses, AT$_{\infty}$ and FD$_{\infty}$, are consistently effective against all attacks except RFA$_{\infty}$, which uses an adversarially-trained surrogate model for generating robust perturbations.
The performance of AT$_{2}$ is also not sensitive to the attack method but sub-optimal due to the use of a different, $L_2$ norm.
In contrast, input pre-processing defenses are generally not useful, although they are known to be effective against white-box attacks where the perturbations are much smaller~\cite{xu2017feature,xie2017mitigating}.
\textbf{Defenses may severely overfit to the type of perturbations that are seen during their optimization.}
First, we notice that NRP is very effective in mitigating smooth perturbations (e.g., generative modeling attacks) but performs much worse against high-frequency perturbations (e.g., gradient stabilization and input augmentation attacks).
This can be explained by the fact that NRP is trained on feature-space, smooth perturbations~\cite{naseer2020self}.
Similarly, AT defenses are not effective in mitigating RFA attacks.
Second, DiffPure is originally claimed to provide state-of-the-art white-box robustness~\cite{nie2022diffusion}.
However, we find that surprisingly, black-box transfer attacks (e.g., RFA and CDA) can effectively bypass DiffPure.
This might be because the Gaussian noise used in the diffusion process of DiffPure is not capable of denoising the semantic perturbations, as shown in Figure~\ref{fig:vis}.
This finding suggests that \textit{DiffPure provides a false sense of security and requires a rightful evaluation following recommendations from}~\cite{athalye2018obfuscated,carlini2019evaluating,tramer2020adaptive}.



\begin{table}[!t]
\caption{Imperceptibility regarding five metrics.
}
\newcommand{\tabincell}[2]{\begin{tabular}{@{}#1@{}}#2\end{tabular}}

\renewcommand{\arraystretch}{1}
      \centering
      \resizebox{\columnwidth}{!}{
        \begin{tabular}{l|cccccc}
\toprule[1pt]
Attacks&PSNR$\uparrow$&SSIM$\uparrow$&$\Delta E$$\downarrow$&LPIPS$\downarrow$&FID$\downarrow$\\
\midrule

PGD&\textbf{28.112}&\underline{0.714}&\textbf{0.662}&\textbf{0.175}&\textbf{33.823}\\
\hline
MI~\cite{dong2018boosting}&24.853&0.576&0.901&0.290&47.570\\
NI~\cite{lin2020nesterov} &26.030&0.619&0.742&0.300&47.637\\
PI~\cite{wang2021boosting}&26.015&0.619&0.741&0.303&50.051\\

\hline
DI~\cite{xie2019improving} &\underline{27.987}&0.713&\underline{0.666}&\underline{0.176}&66.075\\
TI~\cite{dong2019evading} &\underline{27.903}&0.712&0.671&0.178&54.687\\
SI~\cite{lin2020nesterov} &27.712&0.704&0.668&0.187&51.098\\
VT~\cite{wang2021enhancing}&27.264&0.693&0.710&0.200&51.858\\
Admix~\cite{wang2021admix} &27.817&0.703&0.670&0.187&43.052\\

\hline
TAP~\cite{zhou2018transferable} &25.710&0.560&0.743&0.314&84.705\\
AA~\cite{inkawhich2019feature}&26.613&0.665&0.793&0.246&71.301\\
ILA~\cite{huang2019enhancing}&26.222&0.625&0.665&0.223&119.047\\
FIA~\cite{wang2021feature} &26.533&0.647&0.755&0.225&139.522\\
NAA~\cite{zhang2022improving}&26.634&0.653&0.757&0.207&109.383\\

\hline
SGM~\cite{wu2020skip} &27.586&0.704&\underline{0.663}&0.180&\underline{40.637}\\
LinBP~\cite{guo2020backpropagating}&26.884&0.680&0.708&0.196&87.930\\
RFA$_{2}$~\cite{springer2021little} &27.113&\underline{0.720}&0.701&0.185&70.065\\
RFA$_{\infty}$~\cite{springer2021little} &24.542&\textbf{0.751}&0.906&0.226&60.349\\
IAA~\cite{zhu2021rethinking} &26.480&0.671&0.773&0.205&145.231\\
DSM~\cite{yang2022boosting}&27.947&0.712&0.667&\underline{0.177}&\underline{45.909}\\

\hline
GAP~\cite{poursaeed2018generative}&25.184&0.611&0.847&0.285&145.964\\
CDA~\cite{naseer2019cross}&24.200&0.605&1.022&0.337&1098.624\\
GAPF~\cite{kanth2021learning} &25.423&0.641&0.846&0.250&215.241\\
BIA~\cite{zhang2022beyond}  &24.590&0.537&0.883&0.400&230.470\\
TTP~\cite{naseer2021generating} &25.813&0.678&0.764&0.270&209.802\\

\bottomrule[1pt]
\end{tabular}
}
\label{tab:imper}
\end{table}

\subsection{Stealthiness Results}
\label{sec:eva-res-stea}

Table~\ref{tab:imper} reports the imperceptibility results for different attacks following five perceptual metrics.
We make the following observations.
\textbf{All existing transferable attacks indeed sacrifice imperceptibility.} Although all attacks share the same $L_{\infty}$ norm, all transfer attacks lead to more perceptible perturbations than the PGD baseline in almost all cases.
One exception is that RFA$_{\infty}$ achieves the highest SSIM score since it introduces semantics-aligned perturbations rather than disrupting the image structures like DI, as shown in  Figure~\ref{fig:vis} left.
\textbf{Generative modeling attacks are the most perceptible.} In particular, their high FID scores suggest the dramatic distribution shift between their adversarial and original images.
This property can also be confirmed by Figure~\ref{fig:vis} right, where different CDA images actually lead to similar, image-agnostic perturbations.

\begin{figure}[!t]
\centering
\includegraphics[width=\columnwidth]{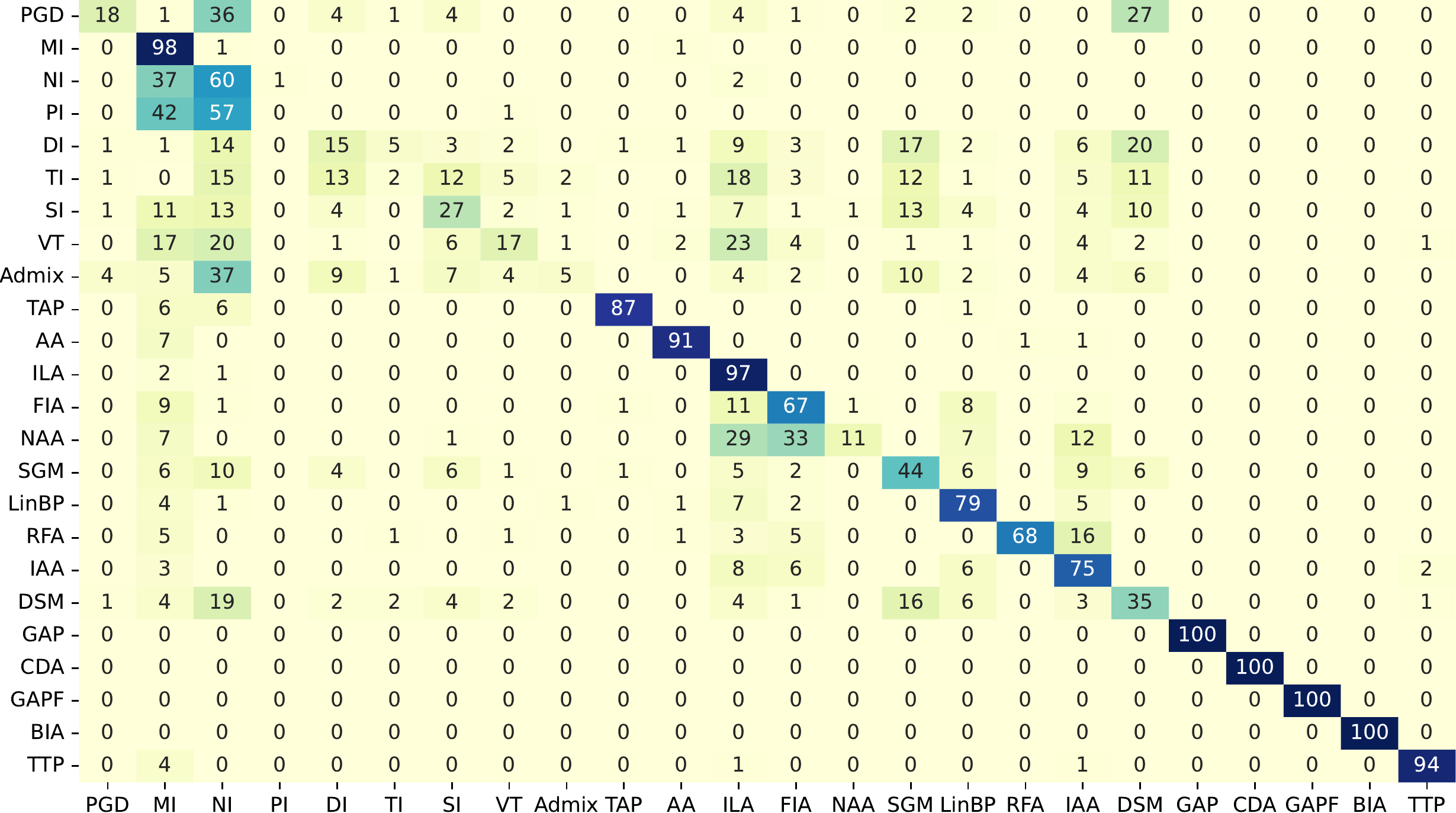}
\caption{Results of attack perturbation classification.
}
\label{fig:confusion}
\end{figure}

In order to shed more light on the perturbation characteristics, we further explore whether different attacks can be simply differentiated based only on their perturbations.
To this end, we perturb 500 randomly selected images using the 24 different attacks, and for each attack, 400 images are used for training a ResNet18 classifier and the rest 100 for testing.
Figure~\ref{fig:confusion} shows that the perturbations generated by different attacks are indeed well separable.
This separability implies that it is possible to trace back the root cause of a successful attack by analyzing its left image perturbations.
More specifically, the fact that attacks in the same category are less separable than the inter-category case confirms that our new attack categorization captures the signature of each specific category.

\begin{figure}[!t]
\centering
\includegraphics[width=\columnwidth,height=3.5cm]{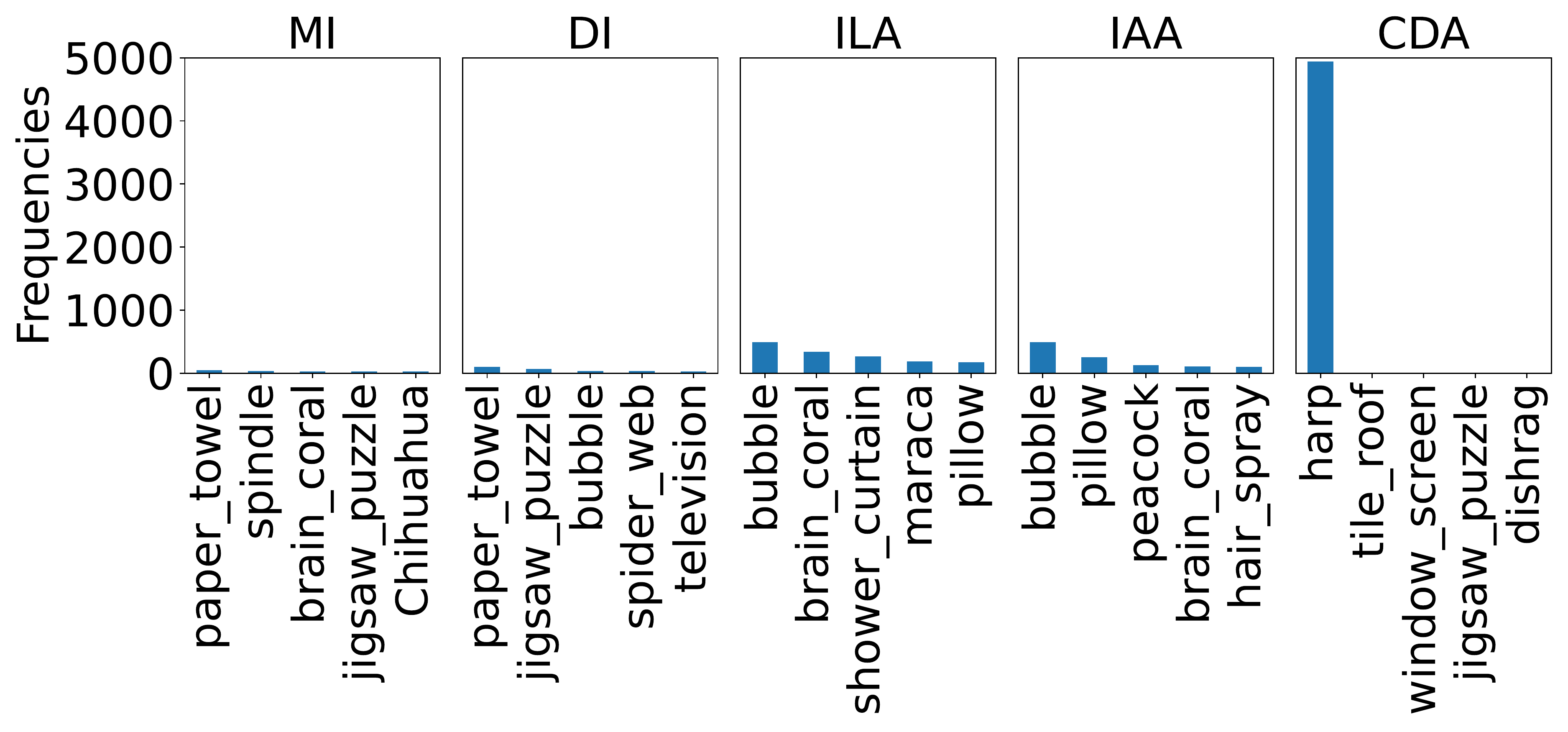}
\caption{Top-5 frequent class predictions over 5000 adversarial images for one attack in each of five categories. Results for all 23 attacks are in Appendix~\ref{app:results}.
}
\label{fig:dis_main}
\end{figure}

In addition to the perturbation characteristics, we also look into the characteristics of misclassification for different attacks.
As can be seen from Figure~\ref{fig:dis_main}, different attack categories lead to different patterns in their resulting model prediction results.
Specifically, gradient stabilization and input augmentation attacks lead to relatively uniform class distributions, and the most frequent class is usually related to semantics that reflect textural patterns, such as ``brain coral''.
In contrast, the other three categories of attacks yield more concentrated class predictions, and especially, generative modeling attacks (e.g., CDA) fool the model to misclassify almost all images into ``harp''.
Such obvious differences also imply the possibility of attack traceback, based on misclassification patterns.

\section{Conclusion}
\label{sec:conc}

In this paper, we have designed good practices in evaluating transfer adversarial attacks.
First, a new attack categorization is proposed to facilitate our systematic and fair analyses of transfer attacks.
The analyses also provide interesting observations that complement or even challenge existing knowledge.
Furthermore, we present a comprehensive evaluation of 23 representative transfer attacks against 9 defenses on ImageNet.
Our extensive evaluation results lead to valuable new insights into both attack transferability and stealthiness.
Our work aims to give a thorough picture of the current progress of transfer attacks, and we hope that it can guide future research towards a more meaningful evaluation of transfer attacks.

\section{Acknowledgements}

This publication has received funding from the Excellence Initiative of Aix-Marseille Université - A*Midex, a French ``Investissements d'Avenir programme'' (AMX-21-IET-017), the UnLIR ANR project (ANR-19-CE23-0009), and the CAS Project for Young Scientists in Basic Research (Grant YSBR-040). Part of this work was performed using HPC resources from GENCI-IDRIS (Grant 2020-AD011013110).

{\small
\bibliographystyle{ieee_fullname}
\bibliography{ref}
}

\clearpage
\newpage

\appendix
\section{General Settings in Existing Work}
\label{app:comp}

\begin{table}[h]
\newcommand{\tabincell}[2]{\begin{tabular}{@{}#1@{}}#2\end{tabular}}
\caption{General settings in existing work regarding five aspects. For each aspect (A/B), \fullcirc~ denotes only considering A, \emptycirc~ denotes only considering B, and \halfcirc~ denotes both.}

\renewcommand{\arraystretch}{1}
      \centering
      \resizebox{\columnwidth}{!}{
        \begin{tabular}{l|ccccc}
\toprule[1pt]
Attacks&\tabincell{c}{Scenario\\(public model/\\private system)}&\tabincell{c}{Attack goal\\ (untarget/\\target)}&\tabincell{c}{Task\\(ImageNet/\\others)}&\tabincell{c}{Model\\ (CNNs/\\others)}&\tabincell{c}{Source/target\\training data\\ (same/\\disjoint)}\\

\midrule
MI~\cite{dong2018boosting}          &\fullcirc&\fullcirc&\fullcirc&\fullcirc&\fullcirc\\
NI~\cite{lin2020nesterov}           &\fullcirc&\fullcirc&\fullcirc&\fullcirc&\fullcirc\\
PI~\cite{wang2021boosting}          &\fullcirc&\fullcirc&\fullcirc&\fullcirc&\fullcirc\\
DI~\cite{xie2019improving}          &\fullcirc&\fullcirc&\fullcirc&\fullcirc&\fullcirc\\
TI~\cite{dong2019evading}           &\fullcirc&\fullcirc&\fullcirc&\fullcirc&\fullcirc\\
SI~\cite{lin2020nesterov}           &\fullcirc&\fullcirc&\fullcirc&\fullcirc&\fullcirc\\ 
VT~\cite{wang2021enhancing}         &\fullcirc&\fullcirc&\fullcirc&\fullcirc&\fullcirc\\ 
Admix~\cite{wang2021admix}          &\fullcirc&\fullcirc&\fullcirc&\fullcirc&\fullcirc\\
TAP~\cite{zhou2018transferable}     &\fullcirc&\fullcirc&\fullcirc&\fullcirc&\fullcirc\\
AA~\cite{inkawhich2019feature}      &\fullcirc&\halfcirc&\fullcirc&\fullcirc&\fullcirc\\
ILA~\cite{huang2019enhancing}       &\fullcirc&\fullcirc&\fullcirc&\fullcirc&\fullcirc\\
FIA~\cite{wang2021feature}          &\fullcirc&\fullcirc&\fullcirc&\fullcirc&\fullcirc\\
NAA~\cite{zhang2022improving}       &\fullcirc&\fullcirc&\fullcirc&\fullcirc&\fullcirc\\
SGM~\cite{wu2020skip}               &\fullcirc&\fullcirc&\fullcirc&\fullcirc&\fullcirc\\
LinBP~\cite{guo2020backpropagating} &\fullcirc&\fullcirc&\fullcirc&\fullcirc&\fullcirc\\
RFA~\cite{springer2021little}       &\fullcirc&\halfcirc&\fullcirc&\halfcirc&\fullcirc\\
IAA~\cite{zhu2021rethinking}        &\halfcirc &\halfcirc&\fullcirc&\fullcirc&\fullcirc\\
DSM~\cite{yang2022boosting}         &\fullcirc&\fullcirc&\halfcirc&\fullcirc&\fullcirc\\
GAP~\cite{poursaeed2018generative}  &\fullcirc&\halfcirc&\halfcirc&\fullcirc&\fullcirc\\
CDA~\cite{naseer2019cross}          &\fullcirc&\fullcirc&\fullcirc&\fullcirc&\halfcirc\\
GAPF~\cite{kanth2021learning}       &\fullcirc&\fullcirc&\fullcirc&\fullcirc&\fullcirc\\
BIA~\cite{zhang2022beyond}          &\fullcirc&\fullcirc&\fullcirc&\fullcirc&\halfcirc\\
TTP~\cite{naseer2021generating}     &\fullcirc&\halfcirc&\fullcirc&\fullcirc&\halfcirc\\
Ours     &\fullcirc&\fullcirc&\fullcirc&\halfcirc&\fullcirc\\

\bottomrule[1pt]
\end{tabular}
}
\label{tab:circle}
\end{table}

\section{Descriptions of Attacks and Defenses}
\label{app:att_def}
\subsection{Descriptions of Attacks}
   \noindent\textbf{Momentum Iterative (MI)}~\cite{dong2018boosting} integrates the momentum term into the iterative optimization of attacks, in order to stabilize the update directions and escape from poor local maxima. This momentum term accumulates a velocity vector in the gradient direction of the loss function across iterations.
   
   \noindent\textbf{Nesterov Iterative (NI)}~\cite{lin2020nesterov} is based on an improved momentum method that further also leverages the look ahead property of Nesterov Accelerated Gradient (NAG) by making a jump in the direction of previously accumulated gradients before computing the gradients in the current iteration. This makes the attack escape from poor local maxima more easily and faster.
   
   \noindent\textbf{Pre-gradient guided Iterative (PI)}~\cite{wang2021boosting} follows a similar idea as NI, but it makes a jump based on only the gradients from the last iteration instead of all previously accumulated gradients.
   
   \noindent\textbf{Diverse Inputs (DI)}~\cite{xie2019improving} applies random image resizing and padding to the input image before calculating the gradients in each iteration of the attack optimization. This approach aims to prevent attack overfitting (to the white-box, source model), inspired by the data augmentation techniques used for preventing model overfitting.
   
   \noindent\textbf{Translation Invariant (TI)}~\cite{dong2019evading} applies random image translations for input augmentation. It also introduces an approximate solution to improve the attack efficiency by directly computing locally smoothed gradients on the original image through the convolution operations rather than computing gradients multiple times for all potential translated images.
   
   \noindent\textbf{Scale Invariant (SI)}~\cite{lin2020nesterov} applies random image scaling for input augmentation. It scales pixels with a factor of $1/2^i$. In particular, in each iteration, it takes an average of gradients on multiple augmented images rather than using only one augmented image as in previous input augmentation attacks.
   
   \noindent\textbf{Variance Tuning (VT)}~\cite{wang2021enhancing} applies uniformly distributed additive noise to images for input augmentation and also calculates average gradients over multiple augmented images in each iteration. 
   
   \noindent\textbf{Adversarial mixup (Admix)}~\cite{wang2021admix} calculates the gradients on a composite image that is made up of the original image and another image randomly selected from an incorrect class. The original image label is still used in the loss function.
   
   \noindent\textbf{Transferable Adversarial Perturbations (TAP)}~\cite{zhou2018transferable} proposes to maximize the distance between original images and their adversarial examples in the intermediate feature space and also introduces a regularization term for reducing the variations of the perturbations and another regularization term with the cross-entropy loss.
   
   \noindent\textbf{Activation Attack (AA)}~\cite{inkawhich2019feature} drives the feature-space representation of the original image towards the representation of a target image that is selected from another class. Specifically, AA can achieve targeted misclassification by selecting a target image from that specific target class. 
   
   \noindent\textbf{Intermediate Level Attack (ILA)}~\cite{huang2019enhancing} optimizes the adversarial examples in two stages, with the cross-entropy loss used in the first stage to determine the initial perturbations, which will be further fine-tuned in the second stage towards larger feature distance while maintaining the initial perturbation directions.
   
   \noindent\textbf{Feature Importance-aware Attack (FIA)}~\cite{wang2021feature} proposes to only disrupts important features. Specifically, it measures the importance of features based on the aggregated gradients with respect to feature maps computed on a batch of transformed original images that are achieved by random image masking.
   
   \noindent\textbf{Neuron Attribution-based Attacks (NAA)}~\cite{zhang2022improving} relies on an advanced neuron attribution method to measure the feature importance more accurately. It also introduces an approximation approach to conducting neuron attribution with largely reduced computations.
   
   \noindent\textbf{Skip Gradient Method (SGM)}~\cite{wu2020skip} suggests using ResNet-like architectures as the source model during creating the adversarial examples. Specifically, it shows that backpropagating gradients through skip connections lead to higher transferability than through the residual modules.
   
   \noindent\textbf{Linear BackPropagation (LinBP)}~\cite{guo2020backpropagating} is proposed based on the new finding that the non-linearity of the commonly-used ReLU activation function substantially limits the transferability. To address this limitation, the ReLU is replaced by a linear function during only the backpropagation process.

   \noindent\textbf{Robust Feature-guided Attack (RFA)}~\cite{springer2021little} proposes to use an adversarially-trained model (with $L_{2}$ or $L_{\infty}$ bound) as the source model based on the assumption that modifying more robust features yields more generalizable (transferable) adversarial examples.
   
   \noindent\textbf{Intrinsic Adversarial Attack (IAA)}~\cite{zhu2021rethinking} finds that disturbing the intrinsic data distribution is the key to generating transferable adversarial examples. Based on this, it optimizes the hyperparameters of the Softplus and the weights of skip connections per layer towards aligned attack directions and data distribution. 
   
     \noindent\textbf{Dark Surrogate Model (DSM)}~\cite{yang2022boosting} is trained from scratch with additional ``dark'' knowledge, which is achieved by training with soft labels from a pre-trained teacher model and using data augmentation techniques, such as Cutout, Mixup, and CutMix. 
     
     \noindent\textbf{Generative Adversarial Perturbations (GAP)}~\cite{poursaeed2018generative} proposes a new attack approach that is based on generative modeling. Specifically, it uses the source classifier as the discriminator and trains a generator using the cross-entropy loss. Once trained, the generator can be used to generate an adversarial example for each input original image with only one forward pass.
     
   \noindent\textbf{Cross-Domain Attack (CDA)}~\cite{naseer2019cross} follows the GAP pipeline but uses a more advanced loss (i.e., relativistic cross entropy) to train the generator. This new loss explicitly enforces the probability gap between the clean and adversarial images, boosting the transferability, especially in cross-domain scenarios.
   
   \noindent\textbf{Transferable Targeted Perturbations (TTP)}~\cite{naseer2021generating} is focused on improving transferability of targeted attacks. It is based on learning target-specific generators, each of which is trained with the objective of matching the distribution of targeted perturbations with that of data from a specific target class. Specifically, input augmentation and smooth perturbation projection are used to further boost the performance.
   
   \noindent\textbf{Generative Adversarial Feature Perturbations (GAFP)}~\cite{kanth2021learning} follows the general pipeline of GAP but trains the generator using a loss that maximizes the feature map distance between adversarial and original images at mid-level CNN layers. 
   
   \noindent\textbf{Beyond ImageNet Attack (BIA)}~\cite{zhang2022beyond} also follows the general pipeline of GAP and specifically introduces a random normalization module to simulate different training data distributions and also a feature distance loss that is only based on important/generalizable features.

\subsection{Descriptions of Defenses}

   \noindent\textbf{Bit-Depth Reduction (BDR)}~\cite{xu2017feature} pre-processes input images by reducing the color depth of each pixel while maintaining the semantics. This operation can eliminate pixel-level adversarial perturbations from adversarial images but have little impact on model predictions of clean images.
   
   \noindent\textbf{Pixel Deflection (PD)}~\cite{prakash2018deflecting} pre-processes input images by randomly replacing some pixels with randomly selected pixels from their local neighborhood. It is specifically designed to happen more frequently to non-salient pixels, and a subsequent wavelet-based denoising operation is used to soften the corruption.
 
   \noindent\textbf{Resizing and Padding (R\&P)}~\cite{xie2017mitigating} pre-processes input images by random resizing, which resizes the input images to a random size, and then random padding, which pads zeros around the resized input images. 
   
   \noindent\textbf{High-level representation Guided Denoiser (HGD)}~\cite{liao2018defense} learns a purification network that can be used to purify/denoise the adversarial perturbations. Specifically, different from previous methods that focus on image-space denoising, HGD minimizes the difference between the clean image and the denoised image at intermediate feature layers. 
   
   \noindent\textbf{Neural Representation Purifier (NRP)}~\cite{naseer2020self} learns a purification network using a combined loss that calculates both the image- and feature-space differences. Specifically, the adversarial images used for training the purification network are generated by feature loss-based adversaries, which are shown to be more effective in handling unseen attacks.
   
   \noindent\textbf{Diffusion Purification (DiffPure)}~\cite{nie2022diffusion} uses a diffusion model as the purification network. It diffuses an input image by gradually adding noise in a forward diffusion process and then recovers the clean image by gradually denoising the image in a reverse generative process.
  The reverse process is shown to be also capable of removing adversarial perturbations.
  
\noindent\textbf{$L_{\infty}$-Adversarial Training (AT$_{\infty}$)}~\cite{xie2019feature} follows the basic pipeline of adversarial training, which is to train the robust model on adversarial images. Specifically, these adversarial images are generated using the PGD targeted attacks with the $L_{\infty}$ distance, and the model is trained distributedly on 128 Nvidia V100 GPUs.

\noindent\textbf{$L_{\infty}$-Adversarial Training with Feature Denoising (FD$_{\infty}$)}~\cite{xie2019feature} modifies the standard model architecture by introducing new building blocks that are designed for denoising feature maps based on non-local means or other filters. This modification can help suppress the potential disruptions caused by the adversarial perturbations, and the modified model is trained end-to-end.

   \noindent\textbf{$L_{2}$-Adversarial Training (AT$_{2}$)}~\cite{salman2020adversarially} trains the model on adversarial images that are generated using the PGD non-targeted attacks with the $L_{2}$ distance.

\subsection{Hyperparameters of Attacks and Defenses}
\label{app:hyper}

\begin{table}[!t]
\newcommand{\tabincell}[2]{\begin{tabular}{@{}#1@{}}#2\end{tabular}}
\caption{Hyperparameters of attacks and defenses.}

\renewcommand{\arraystretch}{1}
      \centering
      \resizebox{\columnwidth}{!}{
        \begin{tabular}{l|c}
\toprule[1pt]
Attacks&Hyperparameter\\

\midrule
MI~\cite{dong2018boosting}&decay factor $\mu=1$\\
NI~\cite{lin2020nesterov} &decay factor $\mu=1$\\
PI~\cite{wang2021boosting} &decay factor $\mu=1$\\
\hline
DI~\cite{xie2019improving} &\tabincell{c}{resize\&pad range $R=[1,~1.1]$\\transformation probability $p=0.7$}\\
TI~\cite{dong2019evading} &translation range $R=[-2,2]$\\
SI~\cite{lin2020nesterov} &scale range $R=[0.1,1]$\\ 
VT~\cite{wang2021enhancing} &noise range $R=[-1.5\epsilon,1.5\epsilon]$\\ 
Admix~\cite{wang2021admix} &mixing factor $\eta=0.2$ \\
\hline
TAP~\cite{zhou2018transferable} & $\lambda = 0.005$, $\eta = 0.01$, $\alpha = 0.5$\\
AA~\cite{inkawhich2019feature}& $20$ random target images from 4 classes\\
ILA~\cite{huang2019enhancing} & ILA projection loss\\
FIA~\cite{wang2021feature}& $N=30$, $P_\mathrm{drop}=0.3$\\
NAA~\cite{zhang2022improving} & $N=30$, $\gamma=1.0$, linear transformation\\
\hline
SGM~\cite{wu2020skip} &decay parameter $\gamma=0.5$\\
LinBP~\cite{guo2020backpropagating} &first residual unit in the third meta block\\
RFA~\cite{springer2021little} & $L_{2}$ $\epsilon=0.1$ or $L_{\infty}$ $\epsilon=8$ PGD-AT\\
IAA~\cite{zhu2021rethinking} &$\beta=15$\\
DSM~\cite{yang2022boosting}&CutMix augmentation\\
\hline
GAP~\cite{poursaeed2018generative}&ResNet152 discriminator\\
CDA~\cite{naseer2019cross} &ResNet152 discriminator\\
GAPF~\cite{kanth2021learning} &ResNet152 discriminator\\
BIA~\cite{zhang2022beyond} &ResNet152 discriminator, RN module\\
TTP~\cite{naseer2021generating}&ResNet50 discriminator\\
\midrule

Defenses&Hyperparameters\\

\midrule
BDR~\cite{xu2017feature}&bit depth $D=2$\\
PD~\cite{prakash2018deflecting} &R-CAM: $k=5$, denoising: $\sigma=0.04$\\
R\&P~\cite{xie2017mitigating} &resize\&pad range $R=[1,~1.1]$\\
\hline
HGD~\cite{liao2018defense} &ResNet152-Wide\\
NRP~\cite{naseer2020self} &ResNet (1.2M parameters) \\
DiffPure~\cite{nie2022diffusion} & noise level $t=150$ WideResNet-50-2\\
\hline
AT$_{\infty}$~\cite{xie2019feature}&\tabincell{c}{$L_{\infty}$ $\epsilon=8$, PGD iters $n=30$, lr=$\alpha=1$}\\
FD$_{\infty}$~\cite{xie2019feature}&\tabincell{c}{$L_{\infty}$ $\epsilon=8$, PGD iters $n=30$, lr=$\alpha=1$}\\
AT$_{2}$~\cite{salman2020adversarially} &\tabincell{c}{$L_{2}$ $\epsilon=3.0$, PGD iters $n=7$, lr=$\alpha=0.5$}\\
\bottomrule[1pt]
\end{tabular}
}
\label{tab:hyperparameters}
\end{table}

\section{Imperceptibility Metrics}
\label{app:metric}
  \noindent\textbf{Peak Signal-to-Noise Ratio (PSNR)} measures the ratio of the maximum possible power of a signal (image $\boldsymbol{x}$) to the noise (perturbations $\boldsymbol{\sigma}$) power.
  It is calculated by:
  \begin{equation}\label{eq:psnr}
  \textrm{PSNR}(\boldsymbol{x},\boldsymbol{\sigma})=10\cdot\log_{10}\frac{\max(\boldsymbol{x})^2}{\textrm{MSE}(\boldsymbol{x},\boldsymbol{\sigma})},
  \end{equation}
where the Mean Squared Error (MSE) measures the average squared difference between the two inputs.

  \noindent\textbf{Structural Similarity Index Measure (SSIM)}~\cite{wang2004image} is used for measuring the perceptual quality of digital images. In our case, it measures the structural similarity between the original image $\boldsymbol{x}$ and adversarial image $\boldsymbol{x}'$, which is considered to be a degraded version of $\boldsymbol{x}$.
  It is calculated by: 
    \begin{equation}\label{eq:ssim}
   \textrm{SSIM}(\boldsymbol{x},\boldsymbol{x}')=\frac{{ ( 2 \mu_{\boldsymbol{x}}\mu_{\boldsymbol{x}'}+ c_1 )+  (2 \sigma_{\boldsymbol{x}\boldsymbol{x}'}+c_2)}} 
{(\mu_{\boldsymbol{x}}^2+\mu_{\boldsymbol{x}'}^2+c_1)(\sigma_{\boldsymbol{x}}^2+\sigma_{\boldsymbol{x}'}^2+c_2)}.
  \end{equation}
Descriptions of $c_1$ and $c_1$ and more technical details can be found in~\cite{wang2004image}.

  \noindent\textbf{$\Delta$\textit{E}}~\cite{luo2001development} refers to the CIE $\Delta E$ standard formula used for measuring the perceptual color distance between two pixels. We follow~\cite{zhao2020towards} to use the latest variant, CIEDE2000, since it is shown to align very well with human perception.
It is calculated in the CIELCH space by:
\begin{equation}
\label{eq:deltae}
\begin{gathered}
\Delta E=\sqrt{(\frac{\Delta L'}{k_LS_L})^2+(\frac{\Delta C'}{k_CS_C})^2+(\frac{\Delta H'}{k_HS_H})^2+\Delta R},\\
\Delta R=R_T(\frac{\Delta C'}{k_CS_C})(\frac{\Delta H'}{k_HS_H}),
\end{gathered}
\end{equation}
where $\Delta L'$, $\Delta C'$, $\Delta H'$ denotes the distance between two pixels in their lightness, chroma and hue channels, respectively. $\Delta R$ is an interactive term between chroma and hue differences.
Detailed definitions and explanations of the weighting functions ($S_L$, $S_C$, $S_H$ and $R_T$) and hyperparameters $k_L$, $k_C$ and $k_H$ can be found in~\cite{luo2001development}.
Further, Image-level perceptual color difference is obtained by computing the $L_{2}$ norm of the above $\Delta E$ for each pixel.

  \noindent\textbf{Learned Perceptual Image Patch Similarity (LPIPS)}~\cite{zhang2018unreasonable} is developed for measuring the perceptual similarity between two images. It computes the cosine distance (in
the channel dimension) between features at each given convolutional layer $l$ and averages the results across spatial dimensions $H\times W$ and layers of a specific network $f$:
  \begin{equation}
\label{eq:lpips}
  \textrm{LPIPS}=\sum_{l}\frac{1}{H_{l}W_{l}}\sum _{h,w} \cos(f^{l}_{hw}(\boldsymbol{x}),f^{l}_{hw}(\boldsymbol{x}')),
  \end{equation}
where AlexNet is adopted as $f$ for computational efficiency.

\noindent\textbf{Frechet Inception Distance (FID)}~\cite{szegedy2016rethinking} is originally used to assess the quality of images generated by a generative model and can also be used to assess the quality of adversarial images.
It compares the distribution of adversarial images to that of original images.
Specifically, the output features from the pool3 layer of an Inception-V3 for original and adversarial images are used to fit two multidimensional Gaussian distributions $\mathcal {N}(\mu,\Sigma )$ and $\mathcal {N}(\mu',\Sigma')$, and the FID score is calculated by:
  \begin{equation}
\label{eq:fid}
  \textrm{FID}(\mathcal {N}(\mu,\Sigma),\mathcal {N}(\mu' ,\Sigma' ))=\|\mu– \mu'\|_2^2 + \textrm{tr}(\Sigma + \Sigma'- 2\sqrt{\Sigma\Sigma'}),
  \end{equation}
where $\textrm{tr}$ denotes the trace of the matrix.

\section{Model Property Metrics}
\label{app:inter}
\newcommand{\relu}{\operatorname{relu}}
\newcommand{\gap}{\operatorname{GAP}}
\newcommand{\up}{\operatorname{up}}

\newcommand{\cam}{\textrm{CAM}}
\newcommand{\gcam}{\textrm{Grad-CAM}}
\newcommand{\scam}{\textrm{Score-CAM}}

\newcommand{\AD}{\operatorname{AD}}
\newcommand{\AI}{\operatorname{AI}}
\newcommand{\OM}{\operatorname{OM}}
\newcommand{\LE}{\operatorname{LE}}
\newcommand{\Fo}{\operatorname{F1}}
\newcommand{\prc}{\operatorname{precision}}
\newcommand{\rec}{\operatorname{recall}}
\newcommand{\BA}{\operatorname{BoxAcc}}
\newcommand{\spg}{\operatorname{SP}}
\newcommand{\epg}{\operatorname{EP}}
\newcommand{\SM}{\operatorname{SM}}
\newcommand{\iou}{\operatorname{IoU}}

\newcommand{\alert}[1]{{\color{red}{#1}}}
\newcommand{\sm}{\scriptsize}
\newcommand{\eq}[1]{(\ref{eq:#1})}

\newcommand{\Th}[1]{\textsc{#1}}
\newcommand{\mr}[2]{\multirow{#1}{*}{#2}}
\newcommand{\mc}[2]{\multicolumn{#1}{c}{#2}}
\newcommand{\tb}[1]{\textbf{#1}}
\newcommand{\ch}{\checkmark}

\newcommand{\red}[1]{{\textcolor{red}{#1}}}
\newcommand{\blue}[1]{{\textcolor{blue}{#1}}}
\newcommand{\green}[1]{{\textcolor{green}{#1}}}
\newcommand{\gray}[1]{{\textcolor{gray}{#1}}}

\newcommand{\tran}{^\top}
\newcommand{\mtran}{^{-\top}}
\newcommand{\zcol}{\mathbf{0}}
\newcommand{\zrow}{\zcol\tran}

\newcommand{\ind}{\mathbb{1}}
\newcommand{\expect}{\mathbb{E}}
\newcommand{\nat}{\mathbb{N}}
\newcommand{\zahl}{\mathbb{Z}}
\newcommand{\real}{\mathbb{R}}
\newcommand{\proj}{\mathbb{P}}
\newcommand{\prob}{\operatorname{P}}
\newcommand{\normal}{\mathcal{N}}

\newcommand{\mif}{\textrm{if}\ }
\newcommand{\other}{\textrm{otherwise}}
\newcommand{\minimize}{\textrm{minimize}\ }
\newcommand{\maximize}{\textrm{maximize}\ }
\newcommand{\st}{\textrm{subject\ to}\ }

\newcommand{\id}{\operatorname{id}}
\newcommand{\const}{\operatorname{const}}
\newcommand{\sgn}{\operatorname{sgn}}
\newcommand{\var}{\operatorname{Var}}
\newcommand{\mean}{\operatorname{mean}}
\newcommand{\trace}{\operatorname{tr}}
\newcommand{\diag}{\operatorname{diag}}
\newcommand{\vect}{\operatorname{vec}}
\newcommand{\cov}{\operatorname{cov}}
\newcommand{\sign}{\operatorname{sign}}
\newcommand{\prj}{\operatorname{proj}}

\newcommand{\softmax}{\operatorname{softmax}}
\newcommand{\clip}{\operatorname{clip}}

\newcommand{\defn}{\mathrel{:=}}
\newcommand{\peq}{\mathrel{+\!=}}
\newcommand{\meq}{\mathrel{-\!=}}

\newcommand{\paren}[1]{\left({#1}\right)}
\newcommand{\mat}[1]{\left[{#1}\right]}
\newcommand{\set}[1]{\left\{{#1}\right\}}
\newcommand{\floor}[1]{\left\lfloor{#1}\right\rfloor}
\newcommand{\ceil}[1]{\left\lceil{#1}\right\rceil}
\newcommand{\inner}[1]{\left\langle{#1}\right\rangle}
\newcommand{\norm}[1]{\left\|{#1}\right\|}
\newcommand{\abs}[1]{\left|{#1}\right|}
\newcommand{\frob}[1]{\norm{#1}_F}
\newcommand{\card}[1]{\left|{#1}\right|\xspace}

\newcommand{\diff}{\mathrm{d}}
\newcommand{\der}[3][]{\frac{\diff^{#1}#2}{\diff#3^{#1}}}
\newcommand{\ider}[3][]{\diff^{#1}#2/\diff#3^{#1}}
\newcommand{\pder}[3][]{\frac{\partial^{#1}{#2}}{\partial{{#3}^{#1}}}}
\newcommand{\ipder}[3][]{\partial^{#1}{#2}/\partial{#3^{#1}}}
\newcommand{\dder}[3]{\frac{\partial^2{#1}}{\partial{#2}\partial{#3}}}

\newcommand{\wb}[1]{\overline{#1}}
\newcommand{\wt}[1]{\widetilde{#1}}

\def\xssp{\hspace{1pt}}
\def\ssp{\hspace{3pt}}
\def\msp{\hspace{5pt}}
\def\lsp{\hspace{12pt}}

\newcommand{\cA}{\mathcal{A}}
\newcommand{\cB}{\mathcal{B}}
\newcommand{\cC}{\mathcal{C}}
\newcommand{\cD}{\mathcal{D}}
\newcommand{\cE}{\mathcal{E}}
\newcommand{\cF}{\mathcal{F}}
\newcommand{\cG}{\mathcal{G}}
\newcommand{\cH}{\mathcal{H}}
\newcommand{\cI}{\mathcal{I}}
\newcommand{\cJ}{\mathcal{J}}
\newcommand{\cK}{\mathcal{K}}
\newcommand{\cL}{\mathcal{L}}
\newcommand{\cM}{\mathcal{M}}
\newcommand{\cN}{\mathcal{N}}
\newcommand{\cO}{\mathcal{O}}
\newcommand{\cP}{\mathcal{P}}
\newcommand{\cQ}{\mathcal{Q}}
\newcommand{\cR}{\mathcal{R}}
\newcommand{\cS}{\mathcal{S}}
\newcommand{\cT}{\mathcal{T}}
\newcommand{\cU}{\mathcal{U}}
\newcommand{\cV}{\mathcal{V}}
\newcommand{\cW}{\mathcal{W}}
\newcommand{\cX}{\mathcal{X}}
\newcommand{\cY}{\mathcal{Y}}
\newcommand{\cZ}{\mathcal{Z}}

\newcommand{\vA}{\mathbf{A}}
\newcommand{\vB}{\mathbf{B}}
\newcommand{\vC}{\mathbf{C}}
\newcommand{\vD}{\mathbf{D}}
\newcommand{\vE}{\mathbf{E}}
\newcommand{\vF}{\mathbf{F}}
\newcommand{\vG}{\mathbf{G}}
\newcommand{\vH}{\mathbf{H}}
\newcommand{\vI}{\mathbf{I}}
\newcommand{\vJ}{\mathbf{J}}
\newcommand{\vK}{\mathbf{K}}
\newcommand{\vL}{\mathbf{L}}
\newcommand{\vM}{\mathbf{M}}
\newcommand{\vN}{\mathbf{N}}
\newcommand{\vO}{\mathbf{O}}
\newcommand{\vP}{\mathbf{P}}
\newcommand{\vQ}{\mathbf{Q}}
\newcommand{\vR}{\mathbf{R}}
\newcommand{\vS}{\mathbf{S}}
\newcommand{\vT}{\mathbf{T}}
\newcommand{\vU}{\mathbf{U}}
\newcommand{\vV}{\mathbf{V}}
\newcommand{\vW}{\mathbf{W}}
\newcommand{\vX}{\mathbf{X}}
\newcommand{\vY}{\mathbf{Y}}
\newcommand{\vZ}{\mathbf{Z}}

\newcommand{\va}{\mathbf{a}}
\newcommand{\vb}{\mathbf{b}}
\newcommand{\vc}{\mathbf{c}}
\newcommand{\vd}{\mathbf{d}}
\newcommand{\ve}{\mathbf{e}}
\newcommand{\vf}{\mathbf{f}}
\newcommand{\vg}{\mathbf{g}}
\newcommand{\vh}{\mathbf{h}}
\newcommand{\vi}{\mathbf{i}}
\newcommand{\vj}{\mathbf{j}}
\newcommand{\vk}{\mathbf{k}}
\newcommand{\vl}{\mathbf{l}}
\newcommand{\vm}{\mathbf{m}}
\newcommand{\vn}{\mathbf{n}}
\newcommand{\vo}{\mathbf{o}}
\newcommand{\vp}{\mathbf{p}}
\newcommand{\vq}{\mathbf{q}}
\newcommand{\vr}{\mathbf{r}}
\newcommand{\Vs}{\mathbf{s}}
\newcommand{\vt}{\mathbf{t}}
\newcommand{\vu}{\mathbf{u}}
\newcommand{\vv}{\mathbf{v}}
\newcommand{\vw}{\mathbf{w}}
\newcommand{\vx}{\mathbf{x}}
\newcommand{\vy}{\mathbf{y}}
\newcommand{\vz}{\mathbf{z}}

\newcommand{\vone}{\mathbf{1}}
\newcommand{\vzero}{\mathbf{0}}

\newcommand{\valpha}{{\boldsymbol{\alpha}}}
\newcommand{\vbeta}{{\boldsymbol{\beta}}}
\newcommand{\vgamma}{{\boldsymbol{\gamma}}}
\newcommand{\vdelta}{{\boldsymbol{\delta}}}
\newcommand{\vepsilon}{{\boldsymbol{\epsilon}}}
\newcommand{\vzeta}{{\boldsymbol{\zeta}}}
\newcommand{\veta}{{\boldsymbol{\eta}}}
\newcommand{\vtheta}{{\boldsymbol{\theta}}}
\newcommand{\viota}{{\boldsymbol{\iota}}}
\newcommand{\vkappa}{{\boldsymbol{\kappa}}}
\newcommand{\vlambda}{{\boldsymbol{\lambda}}}
\newcommand{\vmu}{{\boldsymbol{\mu}}}
\newcommand{\vnu}{{\boldsymbol{\nu}}}
\newcommand{\vxi}{{\boldsymbol{\xi}}}
\newcommand{\vomikron}{{\boldsymbol{\omikron}}}
\newcommand{\vpi}{{\boldsymbol{\pi}}}
\newcommand{\vrho}{{\boldsymbol{\rho}}}
\newcommand{\vsigma}{{\boldsymbol{\sigma}}}
\newcommand{\vtau}{{\boldsymbol{\tau}}}
\newcommand{\vupsilon}{{\boldsymbol{\upsilon}}}
\newcommand{\vphi}{{\boldsymbol{\phi}}}
\newcommand{\vchi}{{\boldsymbol{\chi}}}
\newcommand{\vpsi}{{\boldsymbol{\psi}}}
\newcommand{\vomega}{{\boldsymbol{\omega}}}

\newcommand{\rLambda}{\mathrm{\Lambda}}
\newcommand{\rSigma}{\mathrm{\Sigma}}

\newcommand{\vLambda}{\bm{\rLambda}}
\newcommand{\vSigma}{\bm{\rSigma}}

\makeatletter
\newcommand*\bdot{\mathpalette\bdot@{.7}}
\newcommand*\bdot@[2]{\mathbin{\vcenter{\hbox{\scalebox{#2}{$\m@th#1\bullet$}}}}}
\makeatother

\makeatletter
\DeclareRobustCommand\onedot{\futurelet\@let@token\@onedot}
\def\@onedot{\ifx\@let@token.\else.\null\fi\xspace}

\def\eg{\emph{e.g}\onedot} \def\Eg{\emph{E.g}\onedot}
\def\ie{\emph{i.e}\onedot} \def\Ie{\emph{I.e}\onedot}
\def\cf{\emph{cf}\onedot} \def\Cf{\emph{Cf}\onedot}
\def\etc{\emph{etc}\onedot} \def\vs{\emph{vs}\onedot}
\def\wrt{w.r.t\onedot} \def\dof{d.o.f\onedot} \def\aka{a.k.a\onedot}
\def\etal{\emph{et al}\onedot}
\makeatother

We adopt two metrics, Average Increase and Average Drop, from~\cite{chattopadhay2018grad}.
Let $p^c_i$ and $o^c_i$ be the predicted probability for class $c$ given as input respectively the $i$-th image $\vx_i$ and its masked version, and let $n$ be the number of test images.
Class $c$ is taken as the ground truth.

Average increase (AI) measures the percentage of images where the masked image yields a higher class probability than the original; higher is better:
\begin{equation}
	\AI(\%) \defn \frac{1}{n} \sum_i^n \ind_{p^c_i < o^c_i} \cdot 100
\label{eq:ai}
\end{equation}

Average drop (AD) quantifies how much predictive power, measured as class probability, is lost when only the masked regions of the image are used; lower is better:
\begin{equation}
	\AD(\%) \defn \frac{1}{n} \sum_{i=1}^n \frac{[p^c_i - o^c_i]_+}{p^c_i} \cdot 100.
\label{eq:ad}
\end{equation}

In our paper, we measure AI and the reverse of AD to make sure a higher value represents better interpretability.
We measure the model similarity based on the reverse of Kullback-Leibler divergence of output features as defined in ~\cite{inkawhich2020transferable}: 
\begin{equation}
	KL \defn D_{KL}(softmax(f_s(\vx)[c])||softmax(f(\vx)[c])),
\label{eq:kl}
\end{equation}
where $D_{KL}$ is Kullback-Leibler divergence, and $f(\vx)[c]$ donates the class $c$ logit value of model $f$ given input $\vx$.

\section{Additional Results}
\label{app:results}
\begin{figure}[h]
\centering
\includegraphics[width=\columnwidth]{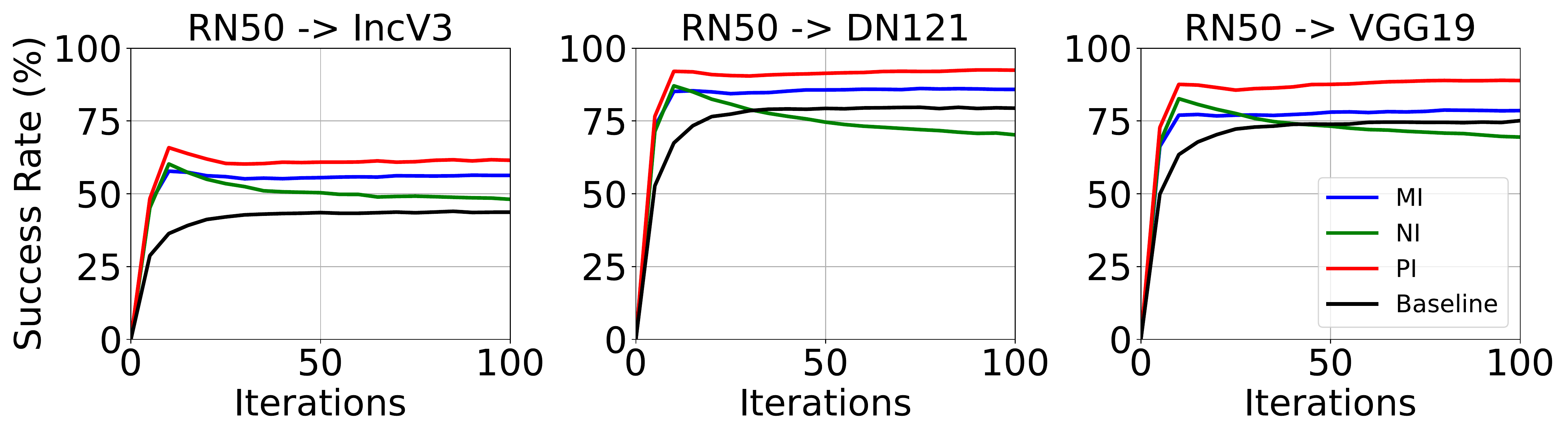}
\caption{Transferability vs. iteration curve on ResNet50 surrogate for gradient stabilization attacks.}
\label{fig:gradient_app}
\end{figure}

\begin{figure}[h]
\centering
\includegraphics[width=\columnwidth]{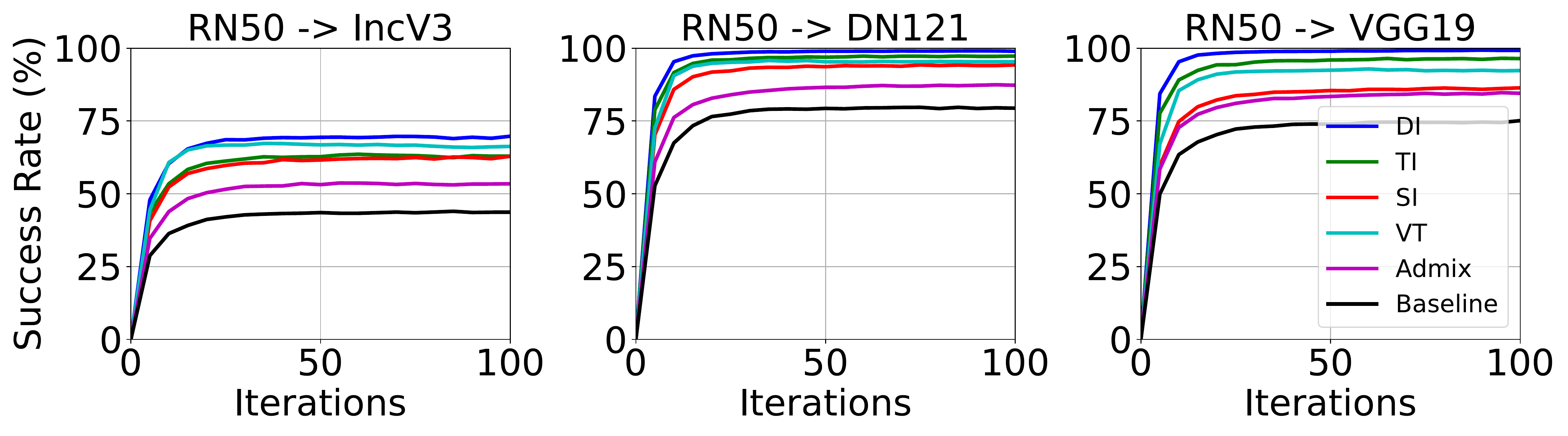}

\caption{Transferability vs. iteration curve on ResNet50 surrogate for input augmentation attacks.}
\label{fig:copy_5_app}
\end{figure}

\begin{figure*}[!t]
\centering
\includegraphics[width=0.9\textwidth,height=5.0cm]{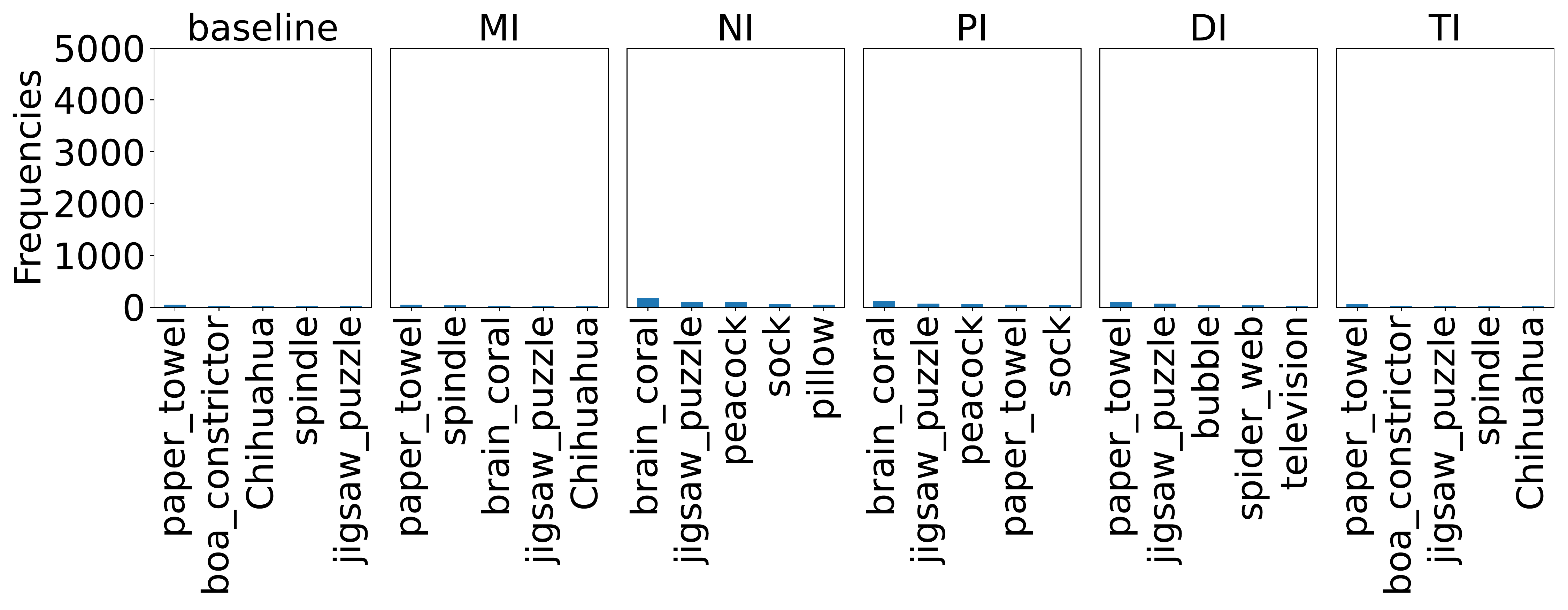}
\includegraphics[width=0.9\textwidth,height=5.0cm]{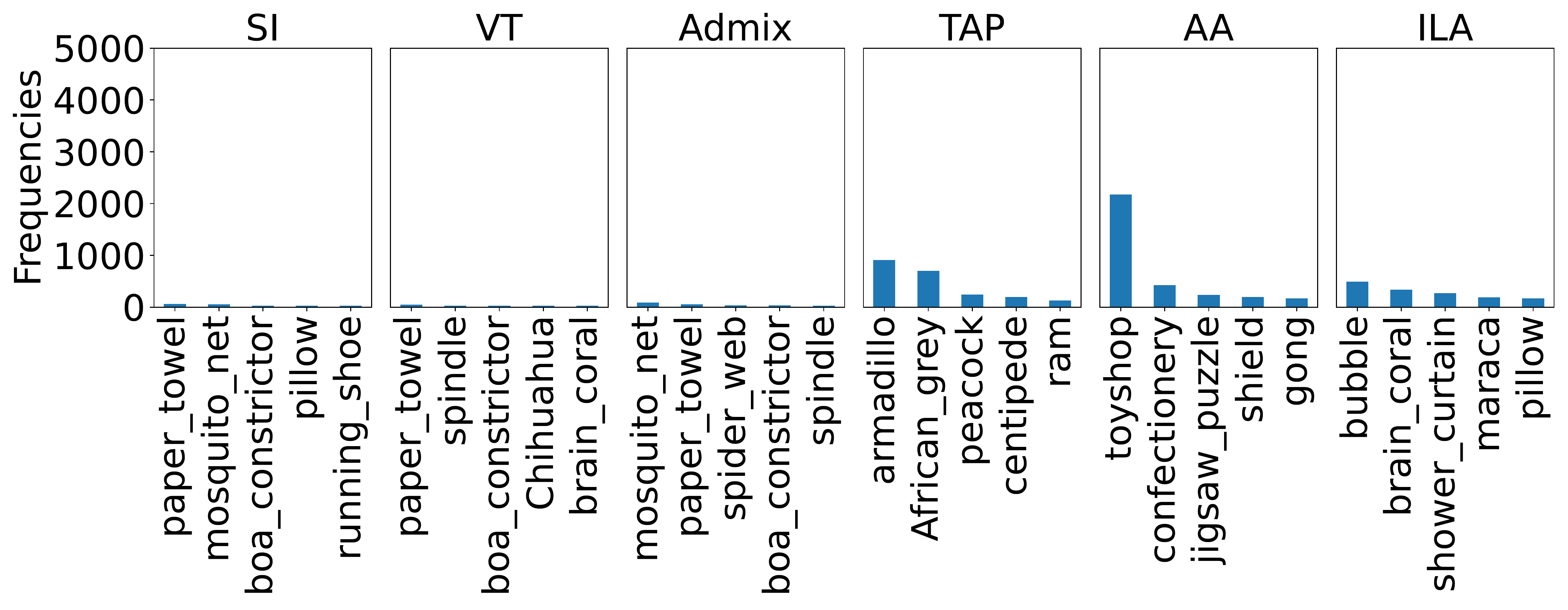}
\includegraphics[width=0.9\textwidth,height=6.1cm]{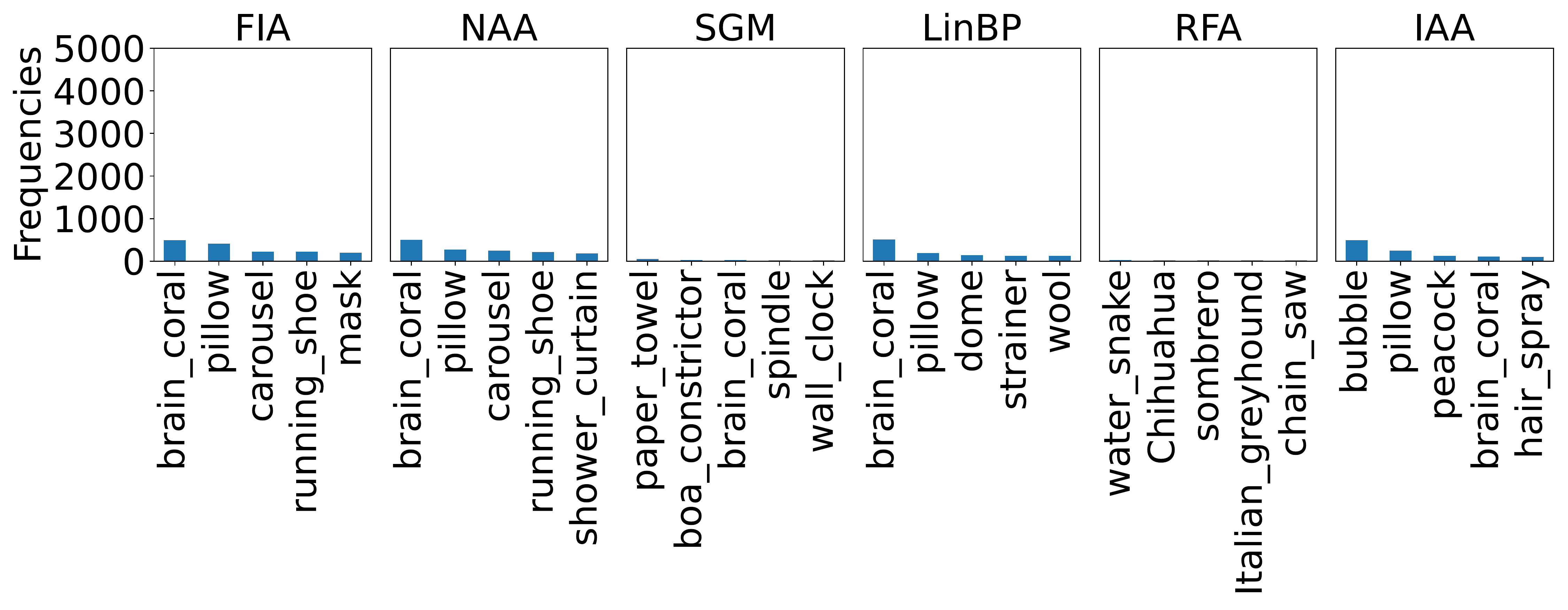}
\includegraphics[width=0.9\textwidth,height=6.0cm]{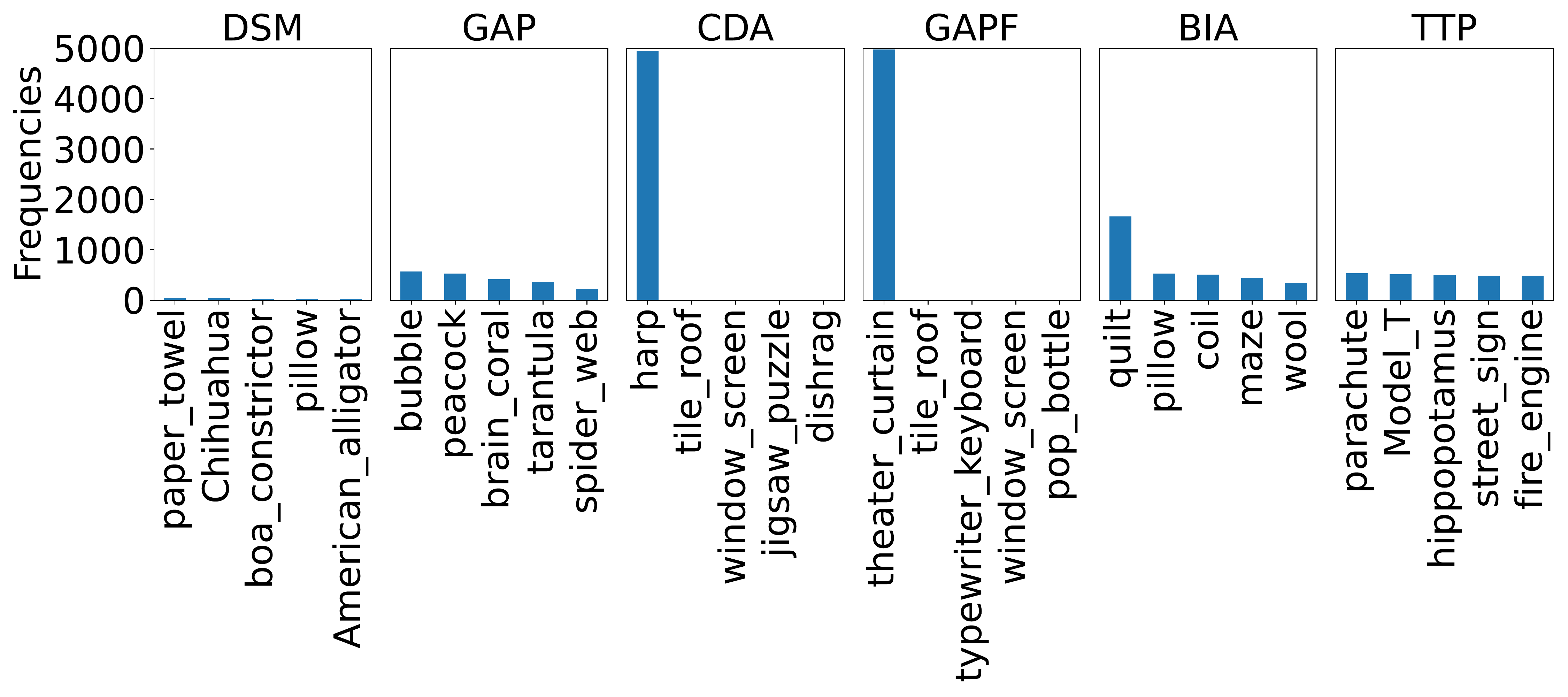}

\caption{Top-5 frequent class predictions calculated over 5000 adversarial images for different attacks.}
\label{fig:dis}
\end{figure*}

\end{document}